# Exact solution to the main turbulence problem for a compressible medium

# and the universal -8/3 law turbulence spectrum of breaking waves

## Sergey G. Chefranov[1)] and Artem S. Chefranov[2)]

[1)] A.M. Obukhov Institute of Atmospheric Physics, Russian Academy of Science, Moscow, Russia;

Russian New University, Moscow, Russia; schefranov@mail.ru

Physics Department, Technion-Israel Institute of Technology, Haifa 32000, Israel; csergei@technion.ac.il

[2)] Ministry of Natural Resources and Environment of the Russia Federation, Moscow, Russia; a.chef@bk.ru

## Abstract

An exact analytical solution to the one-dimensional compressible Euler equations in the form of a nonlinear simple wave is obtained. In contrast to the well-known Riemann solution, the resulting solution and the time of its collapse $t_0$ have an explicit dependence on the initial conditions. For the non-zero dissipation the regularization of the solution over an unlimited time interval is justified. Based on this solution of the Euler equations, an exact explicit and closed description for any single-point and multi-point characteristics of turbulence in a compressible medium are obtained and Onsager's dissipative anomaly is considered. The exact turbulence energy universal spectrum $E(k) \propto k^{-8/3}$, corresponding to the time $t \to t_0$ of the shock arising is stated. That spectrum is more relevant to the strong acoustic turbulence than the well-known spectrum $E(k) \propto k^{-2}$. Installed, spectrum-8/3 is also matched with the observed compressible turbulence spectrum in the magnetosheath and solar wind. The turbulence energy dissipation rate fluctuations universal spectrum $E_D(k) \propto k^{-2/3}$ is obtained and corresponds to the known observation data in the atmospheric surface layer.

# Introduction

Turbulence is an old and unyielding subject for which enormous amounts of data and ideas have been accumulated but for which there are still unsolved problems [1]-[10]. For the sake of certainty, we will proceed from the mathematical formulation of the main problem of the theory of turbulence (T-problem [10]), which is given in [9]. It defines the mathematical basis of the variety of unsolved problems of physics, geophysics, astrophysics, engineering, and even pure mathematics, which are usually associated with the problem of turbulence [7]. According to [9], closed descriptions of any single-point or multi-point correlation moments and spectra of density, pressure and velocity are necessary to obtain the solution of the T-problem. The corresponding hydrodynamic fields must be solutions of the Euler or Navier-Stokes equations in the case of a compressible medium. The very existence of such solutions is still a well-known unsolved problem even for the simpler case (see [9]) where the medium is assumed to be incompressible [11], [12] (www.claymath.org ).

Moreover, the T-problem remains unsolved even for the one-dimensional (1-D) compressible Euler equations despite the fact that the exact solution of the Riemann (1860) problem in the form of a nonlinear simple wave has long been known [13]-[15]. This is due to the fact that the Riemann solution has an implicit form when it is impossible to obtain the dependence of a solution and the value of its finite time of collapse as explicit functions on the arbitrary initial fields. Therefore, until now, the implicit Riemann solution is used in the form of Riemann solver mainly in connection with the numerical simulations [16].

Thus, there is currently no exact solution to the above-mentioned T- problem, regardless of whether the compressibility of the medium is taken into account or not. That is why the problems of describing turbulence in a real compressible medium and small-scale intermittency of turbulence [9], [13], as well as the Onsager (1949) dissipative anomaly problem [1], [7], [17], are unsolved in the modern theory of turbulence.

Indeed, the classical theory of turbulence is formulated for an incompressible medium [9], [13], [17] and for a compressible medium it is problematic its application. So far, only the results of numerical simulation are available, although approximate approaches are being developed to formulate analogs of the laws of this theory [18] - [21].

The classical theory of turbulence is based on the Kolmogorov-Obukhov heuristic law -5/3 for the energy spectrum $E(k) \propto \langle \varepsilon \rangle^{2/3} k^{-5/3}$ and the Kolmogorov law 4/5 for the third-order structural function $S_3 = -\frac{4}{5} \langle \varepsilon \rangle r$, obtained from the Karman-Howarth equation for the second-order structural function $S_2$ of the solenoidal velocity field $\vec{V}$ [9], [13]. Here $\langle \varepsilon \rangle = \nu \left\langle \left( \vec{\nabla} \otimes \vec{V} \right)^2 \right\rangle$ - is the statistically average rate of the kinetic energy viscous dissipation per unit mass. In this case, the value $\langle \varepsilon \rangle$ is assumed to be constant and not equal to zero even in the limit of zero viscosity $\nu \to 0$, which gives a well-known Onsager's problem of the dissipative anomaly in the turbulence theory (see [17]). It consists in the fact that the kinetic energy of turbulence may not be preserved even in the limit of zero viscosity due to the violation of the smoothness of the velocity field in the hypothetical occurrence of a singularity in the solution of the Euler equations [17]. In this connection, for the law 4/5 in [7], a physical anomaly is also noted, associated with the violation of the symmetry to the reversal of the time sign $t \to -t$ because $S_3 \to -S_3$, but $\langle \varepsilon \rangle \to \langle \varepsilon \rangle$.

Finally, the problem of intermittency of small-scale turbulence also remains solved so far only in the framework of semi-empirical models [9], [13], [17]. In this case, for the law -5/3, corrections are introduced related to the heuristic description of the spectra of fluctuations in the energy dissipation rate $\varepsilon$, observed in the experiment, when $E(k) \propto \langle \varepsilon \rangle^{2/3} k^{-\frac{5}{3}} (Lk)^{-q}$, $q = \frac{\mu}{9}$ (see (25.31) in [9] or.(58) in [17]), where $L$ is the integral turbulence scale, which characterizes the turbulence energy pumping scale.

The parameter $\mu$, provided $0 < \mu < 1$, was first introduced in the Novikov-Stewart paper (1964) [22]-[25], where an approach to the problem of small-scale intermittency of turbulence was formulated based on the representations of scale similarity and scale invariance. Similar ideas were independently developed in the well-known works of Mandelbrot (1965) [26] and Kadanov (1966) [27], which served to create the modern theory of fractals and the modern theory of critical phenomena [28], respectively.

Due to the lack of a solution to the T-problem, the concrete value of the parameter $\mu$ is determined so far only on the basis of an experimentally established exponent of the spectrum of fluctuations in the energy dissipation rate $E_D(k) \propto k^{1-\mu}$, corresponding to a structural function of the form $S_D(r) = \left\langle \left( \varepsilon(\vec{x}+\vec{r};t) - \varepsilon(\vec{x};t) \right)^2 \right\rangle \propto r^{-\mu}$ [9].

Measurements in the drive layer of the atmosphere gives $\mu \approx 0.38 \pm 0.05$ [29] (see also Figure 103 in Volume 2 [9] and a similar result in [30]). With subsequent more accurate measurements in the small-scale region of the inertial sub-range, an average value $\mu \approx 0.33$ was obtained for measuring the small-scale intermittency of turbulence in the surface layer of the atmosphere [31].

In our paper, a solution to the T- problem is given based on the exact explicit solution for 1-D compressible Euler equations for arbitrary smooth initial velocity fields that have a finite energy integral in an unbounded space. An averaging over the spatial coordinate is used instead of considering statistical averaging.

At the Euler equations solution collapse condition $t \to t_0$ the exact universal solution to the dissipation spectrum $E_D(k) \propto k^{-2/3}$ with the value $\mu = 1/3$, corresponding to the experimental value $\mu \approx 0.33$ of Kholmyansky (1972) [31] is obtained.

A generalization of Kolmogorov's law 4/5 without any physical anomaly is obtained for the case of turbulence in a compressible medium, since the time reversal symmetry for a non-zero third-order structural function is preserved even at the moment of the solution collapse.

In the region of the singularity of the explicit analytical solution of the Euler equations a hypothetical dissipative anomaly predicted in the Onsager turbulence theory is validated for more general cases than in Onsager's theory. Indeed, according to Onzager's theory [17], and its refinement in [32], a necessary condition for the realization of a dissipative anomaly is a restriction $h \le 1/3$, on the Holder parameter, where the parameter $h$ characterizes the smoothness of the velocity field $|V(x+r) - V(x)| \propto r^h$. The example of the implementation of the dissipative anomaly for the exact solution of the Euler equations given in our paper corresponds to the Holder exponent $h = 5/6$, which does not satisfy the specified restriction of the Onsager theory [17], [32]. However, in [33] a criterion for the realization of a dissipative anomaly is obtained in the form of a condition that corresponds to an inequality $h \leq 5/6$ that agrees with the conclusion obtained in our paper. In contrast to [32], where periodic boundary condition is used, in [33] also as in our paper, considers the case of an unbounded space with a finite integral of the square of the velocity.

The exact solution for the universal spectrum $E(k) \propto k^{-8/3}$ of turbulence energy in a compressible medium, which is independent of the initial velocity field and the adiabatic index of the medium, is obtained for the collapse of the solution at $t \to t_0$. In [33] the threshold spectrum -8/3 law, which is corresponded the Holder index $h = 5/6$ is also considered. The problem of determining the conditions for the implementation of the -8/3 law, which differs significantly from the -5/3 law, is noted in [33] and discussed further in connection with the known data of observations of turbulence spectra in cosmic plasma.

To the one-dimensional Hopf equation (the Burgers equation with zero viscosity) near the collapse of its solution, the energy spectrum with the exponent -8/3 is also obtained in [34], but only for the special case of the sinusoidal initial velocity field and periodic boundary conditions. In this paper, this conclusion of [34] is generalized on the basis of an exact solution of the Hopf equation in Euler variables [2]-[6], for the case of an arbitrary smooth initial velocity field in unbounded space. As a result, it turns out that despite the difference between the solution of the Euler equations and the solution of the Hopf equation, at the moment of collapse of each of these solutions, the energy spectrum has the same form. This indicates that at the time of the collapse of the solution, the role of the effects associated with the pressure gradient in the equations of hydrodynamics may be insignificant. Thus, the spectrum $E(k) \propto k^{-8/3}$ is more relevant to theory [35], than energy spectrum $E(k) \propto k^{-2}$ mentioned in papers [35]-[37]. Indeed, the energy spectrum of acoustic turbulence must correspond exactly to the solution of the Hopf equation according theory [35].

Spectrum with an exponent of -8/3 is also obtained in [38] in the numerical simulation for a distribution of bubble sizes generated by breaking waves.

In [39] (see also [40]), a spectrum with an exponent of -8/3 is also obtained based on the numerical analysis of non-steady self-similar solutions of the second kind for the nonlinear diffusion model of weak turbulence of kinetic Alfven waves. However, after the collapse of the solution, this spectrum is replaced by the well-known law -5/2. Indeed, it is noted in [39] that to date there is no explanation of the spectrum observed for solar wind turbulence with an index of -8/3 within the framework of strong turbulence (-7/3) and weak wave turbulence (-5/2) and this problem is one of the most important in space plasma physics. Energy spectrum -8/3 law is consistent with observational data for the turbulence spectrum in the solar wind and magnetosheaths of the Earth and of Saturn [42]-[47]. According to this observational data, it turns out that the universal turbulence spectrum described by the Kolmogorov-Obukhov law $k^{-5/3}$ changes to a more abrupt mode of decreasing $k^{-8/3}$ long before the characteristic scales of an ordinary dissipative processes [42]-[47]. At the same time, for the magnetolayer of the Earth and Saturn, in contrast to the solar wind, the spectrum with an index of -8/3 can dominate, when the area with a spectrum of -5/3 can be either very insignificant (see Fig.2 below), or none at all [45]. In this case, the problem of determining the mechanism of the appearance of an additional sink of turbulent energy, which leads to the observed fracture in the turbulence spectrum, has not yet been solved. Understanding how collisionless plasmas dissipate energy remains a topic of central importance in space physics, astrophysics, and laboratory plasma studies [46].

The present results provide a step toward understanding this problem on the basis of the exact solution for compressible Euler's equations. It is shown that such energy dissipation in the cosmic plasma can occur due to the collapse of nonlinear mirror modes [48], [49], as is the case for the simple waves considered in this paper.

In Section 1, the well-known procedure (see [13]) is shown for reducing one-dimensional Euler's equations to the Riemann problem is given and the possibility of its further conversion to the Hopf equation for the sum or difference for the flow velocity field and the local speed of sound.

In Section 2, an explicit form of solving the Riemann problem is given based on the known [2]-[6] explicit solution of the Hopf equation in Euler variables. In Section 3, we obtain an exact explicit analytical solution for the Riemann invariants in the special case of isentropic flow with adiabatic index $\gamma = 3$. In this case, an exact explicit analytical solution of the Euler equations is obtained not only for the case of a simple wave, but also for a broader class of isentropic flows with the specified adiabatic exponent. In Section 4, an explicit representation for the time $t_0$ of the shock wave occurrence when the solution collapses is obtained.

In Section 5, the possibility of the shock discontinuity regularization is shown, taking into account a threshold linear homogeneous dissipation or by the introduction of any extremely small viscosity, modeling by the Gaussian delta-correlated in time stochastic velocity field. In Section 6, the exact solution to the T-problem for the correlation functions and corresponding turbulence spectra is obtained and Onsager's dissipative anomaly is considered. In Section 7, a discussion and comparison with observational data and other theories is provided.

## 1. Euler's equations and Riemann solution

Let us show the main steps in the inference of a Riemann solution in the form of a simple wave; these are formulated in [13] and also used here to obtain an exact explicit closed-form analytical solution of this problem.

The one-dimensional Euler equation and the continuity equation are represented in the form [13]:

$$\frac{\partial V}{\partial t} + V \frac{\partial V}{\partial x} + \frac{1}{\rho} \frac{\partial p}{\partial x} = 0 \qquad (1.1)$$

$$\frac{\partial \rho}{\partial t} + \frac{\partial (\rho V)}{\partial x} = 0 \qquad (1.2)$$

For example, to the polytropic medium, in addition to (1.1), (1.2) also the following relation is considered:

$$p = p_{00}(\frac{\rho}{\rho_{00}})^{\gamma} \qquad (1.3)$$

Assuming for the case of simple wave that $V$ can be also represented as a function of $\rho$ only, the following relations are obtained in [13] (see (101.4) in [13]):

$$\frac{dV}{d\rho} = \pm\frac{c}{\rho};$$

$$c^2 = \frac{dp}{d\rho} \tag{1.4}$$

Here, the following forms are used $\frac{1}{\rho}\frac{\partial p}{\partial x} = \frac{1}{\rho}\frac{dp}{dV}\frac{\partial V}{\partial x}$ and

$\frac{\partial(\rho V)}{\partial x} = (V + \rho\frac{dV}{d\rho})\frac{\partial\rho}{\partial x}$ . It allows the derivation in [13] of (1.4) from $(\frac{\partial x}{\partial t})_\rho = (\frac{\partial x}{\partial t})_V$ , where

$(\frac{\partial x}{\partial t})_\rho = -\frac{\partial\rho/\partial t}{\partial\rho/\partial x} = V + \rho\frac{dV}{d\rho}; (\frac{\partial x}{\partial t})_V = -\frac{\partial V/\partial t}{\partial V/\partial x} = V + \frac{1}{\rho}\frac{dp}{dV}$ .

In the results from (1.1), (1.2) and (1.4), the following modification of the one-dimensional hydrodynamic equations can be inferred [13, 14]:

$$\frac{\partial V}{\partial t} + (V \pm c)\frac{\partial V}{\partial x} = 0, \tag{1.5}$$

$$\frac{\partial\rho}{\partial t} + (V \pm c)\frac{\partial\rho}{\partial x} = 0, \tag{1.6}$$

$$\frac{\partial c}{\partial t} + (V \pm c)\frac{\partial c}{\partial x} = 0 \tag{1.7}$$

$$\frac{\partial p}{\partial t} + (V \pm c)\frac{\partial p}{\partial x} = 0 \tag{1.8}$$

In the system of equations (1.5)-(1.8), the plus sign corresponds to the propagation of a simple nonlinear wave described by these equations in the positive direction of the axis $x$, and the minus sign corresponds to its propagation in the opposite direction. In (1.5)-(1.8), the mutual dependence of the fields indicated above will be reflected in the presence of corresponding dependencies of the initial fields for any (not only (1.3)) equation of state of the medium that specifies the type of such dependence.

The definition of the function $c(V)$ used in equations (1.5)-(1.8) is given in (1.4), and that function in [13] is considered as a local sound speed in a non-equilibrium medium moving with the velocity $V(x,t)$.

The first two equations, (1.5) and (1.6), are inferred in [13] from (1.1) and (1.2) directly using (1.4). Equations (1.7) and (1.8) are inferred also taking into account (1.4), (1.5), and (1.6) (see also Formula (1.51) in [14] and the representation following (1.51) for the local speed of sound, which is equivalent to the above equation (1.7)).

From (1.5) and (1.7), in particular, adding and subtracting one equation from another gives the Hopf equation [2-6]:

$$\frac{\partial u}{\partial t} + u\frac{\partial u}{\partial x} = 0;$$
$$u = V \pm c \qquad\qquad (1.9)$$

In Equation (1.9), the case when $u = V + c$ corresponds to a simple wave propagating in the positive direction of the x-axis. The propagation speed of a given point in the wave profile becomes greater with the increasing of density because $du/d\rho > 0$ (see formula (99.10) in [13]). This leads, ultimately, to the breaking of a simple wave and the appearance of a discontinuity at the surface, when the wave ceases to be simple and a shock wave occurs [13]. This distinguishes the nonlinear wave described in (1.9) from linear acoustic waves of small amplitudes propagating at the constant speed of sound.

For the Equations (1.5) and (1.7), implicit solutions are known [14] (see also after (1.51) in [14]):

$$V(x,t) = V_0(x - t(V(x,t) \pm c(x,t))), \qquad (1.10)$$

$$c(x,t) = c_0(x - t(V(x,t) \pm c(x,t))) \qquad (1.11)$$

For example, the determination of the value $t = t_0$ for the shock wave arising time is given in [13], for the case of the equation of state (1.3), when the following form of the Riemann solution is used (see (101.6)-(101.9) in [13]):

$$c(x,t) = c_{V=0} \pm \frac{\gamma - 1}{2}V(x,t);$$
$$\rho(x,t) = \rho_{V=0}(1 \pm \frac{\gamma - 1}{2}\frac{V(x,t)}{c_{V=0}})^{\frac{2}{\gamma-1}}; \qquad (1.12)$$
$$p(x,t) = p_{V=0}(1 \pm \frac{\gamma - 1}{2}\frac{V(x,t)}{c_{V=0}})^{\frac{2\gamma}{\gamma-1}}$$

$$V(x,t) = V_0(x - t(\pm c_{V=0} + \frac{\gamma+1}{2} V(x,t))) \qquad (1.13)$$

In (1.12) the solution of the hydrodynamic equations for the fields of density, pressure and local speed of sound is expressed in terms of one independent velocity field (1.13). In (1.12), hydrodynamic fields of local sound speed, $c(x,t)$, density, $\rho$, pressure, $p$, are expressed via velocity field $V(x,t)$;

$c_{V=0} \equiv c_{V=0}(x,t); \rho_{V=0} \equiv \rho_{V=0}(x,t); p_{V=0} \equiv p_{V=0}(x,t)$, defined for zero velocity under condition "that, in the wave, there is a point where $V = 0 \ldots$"[13]. However, the velocity field $V(x,t)$ itself has no explicit form since it is defined in terms of the solution of the functional equation (1.13), in which the unknown function $V(x,t)$ [13] in the left-hand side of (1.13) appears also in the argument of the arbitrary initial function.

## 2. Explicit solution of the Riemann problem

1. Let us get from (1.10) an explicit form for the velocity field using the approach developed in [2-6].

For that we write (1.10) in the equivalent form:

$$V(x,t) = \int_{-\infty}^{\infty} d\xi V_0(\xi)\delta(\xi - x + tu(x,t)); u(x,t) \equiv V(x,t) \pm c(x,t) \qquad (2.1)$$

In (2.1) $\delta$ is the Dirac delta-function. As noted above, the function $u(x,t)$, used in (2.1), satisfies the 1-D Hopf equation (1.9).

Based on (1.9) and using the known features of delta-function, it is possible to prove the following identity (details of the proof are given in [5] and [6]):

$$\delta(\xi - x + tu(x,t)) \equiv \delta(\xi - x + tu_0(\xi))(1 + t\frac{du_0(\xi)}{d\xi}) \qquad (2.2)$$

In the result, by substitution of (2.2) into (2.1), we get a closed description of an explicit velocity field via the initial velocity field and local sound speed:

$$V(x,t) = \int_{-\infty}^{\infty} d\xi V_0(\xi)(1 + t\frac{du_0(\xi)}{d\xi})\delta(\xi - x + tu_0(\xi)) , \qquad (2.3)$$

$$u_0(\xi) = V_0(\xi) \pm c_0(\xi);$$

For the case when $c_0 = 0$ the solution (2.3) is reduced to the known solution of the Hopf equation which is obtained in [2], [3] (see [4]-[6] where a generalization on 2-D and 3-D cases is also obtained).

Thus, representation (2.3) gives a generalization of known solutions [2-6] to the case of the 1-D Euler equation with nonzero pressure gradient when $c_0 \neq 0$ in (2.3). Solution (2.3) gives the explicit analytical form of the Riemann solution for the simple nonlinear wave.

Note also, that from (1.11) we get a similar relation for the local velocity of sound:

$$c(x,t) = \int_{-\infty}^{\infty} d\xi c_0(\xi)(1 + t\frac{du_0}{d\xi})\delta(\xi - x + tu_0(\xi)) \qquad (2.4)$$

For the distributed density and pressure, we also have the following exact solutions of the equations (1.6) and (1.8), respectively:

$$\rho(x,t) = \int_{-\infty}^{\infty} d\xi \rho_0(\xi)(1 + t\frac{du_0}{d\xi})\delta(\xi - x + tu_0(\xi)); \rho_0(x) = \rho(x,t = 0) \qquad (2.5)$$

$$p(x,t) = \int_{-\infty}^{\infty} d\xi p_0(\xi)(1 + t\frac{du_0}{d\xi})\delta(\xi - x + tu_0(\xi)); p_0(x) = p(x,t = 0) \qquad (2.6)$$

The solutions obtained above are true for an arbitrary medium, not only for a polytropic gas (1.3), because relations (2.3)-(2.6) do not use any particular equation of state defining a pressure-density relation.

2. Let us verify now, for example, that the density field (2.5) actually exactly satisfies (1.6) if velocity and local sound speed fields are represented by (2.3) and (2.4), respectively. For that, substitute (2.5) into (1.6), taking into account (2.3)and (2.4). Also, as in [4]-[6] we used identities which are as follows:

$$\frac{\partial\delta(\xi - x + tu_0(\xi))}{\partial x} = -\frac{1}{(1 + t\frac{du_0(\xi)}{d\xi})}\frac{\partial\delta(\xi - x + tu_0(\xi))}{\partial\xi} \qquad (2.7)$$

$$\delta(\xi_1 - \xi + t(u_0(\xi_1) - u_0(\xi))) = \frac{\delta(\xi_1 - \xi)}{(1 + t\frac{du_0(\xi)}{d\xi})} \tag{2.8}$$

Then, for the first term in (1.6), we have by using (2.5):

$$A \equiv \frac{\partial \rho}{\partial t} = \int_{-\infty}^{\infty} d\xi \rho_0(\xi) \left[ \frac{du_0}{d\xi} \delta(\xi - x + t u_0(\xi)) - u_0(\xi)(1 + t\frac{du_0}{d\xi}) \frac{\partial \delta(\xi - x + t u_0(\xi))}{\partial x} \right] \tag{2.9}$$

Accounting for (2.7), from (2.9) we get after integrating by parts:

$$A = -\int_{-\infty}^{\infty} d\xi u_0(\xi) \frac{d\rho_0}{d\xi} \delta(\xi - x + t u_0(\xi)) \tag{2.10}$$

For the second term in (1.6), by using (2.5) and (2.3), (2.4), we get:

$$B \equiv (V \pm c)\frac{\partial \rho}{\partial x} = \int_{-\infty}^{\infty} d\xi \int_{-\infty}^{\infty} d\xi_1 \rho_0(\xi) u_0(\xi_1)(1 + t\frac{du_0}{d\xi_1})\delta(\xi_1 - x + t u_0(\xi_1))(1 + t\frac{du_0}{d\xi})\frac{\partial \delta(\xi - x + t u_0(\xi))}{\partial x} \tag{2.11}$$

Applying (2.7), from (2.11), we have:

$$B = \int_{-\infty}^{\infty} d\xi \int_{-\infty}^{\infty} d\xi_1 \frac{d\rho}{d\xi} u_0(\xi_1)(1 + t\frac{du_0}{d\xi_1})\delta(\xi_1 - x + t u_0(\xi_1))\delta(\xi - x + t u_0(\xi)) \tag{2.12}$$

Using a delta-function feature, we get for the product of these functions in (2.12):

$$\delta(\xi_1 - x + t u_0(\xi_1))\delta(\xi - x + t u_0(\xi)) = \delta(\xi_1 - \xi + t(u_0(\xi_1) - u_0(\xi)))\delta(\xi - x + t u_0(\xi)) \tag{2.13}$$

From (2.12), accounting for (2.13) and (2.8) and integrating by parts over the variable $\xi_1$, we get that $B = -A$,

where $A$ is represented in (2.10).

Thus, it is proved that the sum of the first two terms in (1.6) $A + B \equiv 0$ identically equals to zero.

3. Let us also consider the partial derivative of density in respect to the space variable. From (2.5) and (2.7) it is possible to obtain:

$$\frac{\partial \rho(x,t)}{\partial x} = \int_{-\infty}^{\infty} d\xi \frac{d\rho_0(\xi)}{d\xi} \delta(\xi - x + t u_0(\xi)) \tag{2.13}$$

From (2.13) and (2.8) for the density gradient square we get a representation in the explicit closed form:

$$\left(\frac{\partial \rho}{\partial x}\right)^2 = \int\limits_{-\infty}^{\infty} d\xi \frac{(d\rho_0/d\xi)^2 \delta(\xi - x + tu_0(\xi))}{1 + tdu_0/d\xi} \tag{2.14}$$

Let us consider the example of a polytropic gas with the isentropic equation of state (1.3), when the initial velocity field has the arbitrary form $V_0(x)$ in (2.14). In this case $V_0(x) \to 0$ for $x \to \infty$, when in (2.14) according to (1.12) we have the following representations for the initial distributions of the density and local speed of sound in terms of the initial distribution of the velocity field:

$$\rho_0(x) = \rho_\infty \left(1 \pm \frac{(\gamma - 1)}{2c_\infty} V_0(x)\right)^{\frac{2}{\gamma - 1}} \tag{2.15}$$

$$c_0(x) = c_\infty \pm \frac{(\gamma - 1)}{2} V_0(x);$$

$$p_0(x) = p_\infty \left(1 \pm \frac{(\gamma - 1)}{2c_\infty} V_0(x)\right)^{\frac{2\gamma}{\gamma - 1}} \tag{2.16}$$

In (2.15) and (2.16) $\rho_\infty = const; c_\infty^2 = \frac{p_\infty \gamma}{\rho_\infty} = const$ are the constant equilibrium values of the density and local speed of sound at an infinite distance from the origin, where the initial perturbation of the velocity field tends to zero. Taking into account (2.16), from (2.3) and (2.6) we obtain an exact explicit analytical solution of the 1-D Euler equations for the velocity and pressure fields:

$$V(x,t) = \int\limits_{-\infty}^{\infty} d\xi V_0(\xi) \left(1 + t\frac{(\gamma + 1)}{2}\frac{dV_0}{d\xi}\right) \delta\left(\xi - x + t\left(\pm c_\infty + \frac{(\gamma + 1)}{2} V_0(\xi)\right)\right) \tag{2.17}$$

$$p(x;t) = \int\limits_{-\infty}^{\infty} d\xi p_0(\xi) \left(1 + t\frac{(\gamma + 1)}{2}\frac{dV_0}{d\xi}\right) \delta\left(\xi - x + t\left(\pm c_\infty + \frac{(\gamma + 1)}{2} V_0(\xi)\right)\right) \tag{2.18}$$

When considering the formal limit $c_\infty \to 0$ and $\gamma \to 1$ in (2.17), when the representation of (2.17) for the velocity field exactly coincides with the known solution of the Hopf equation in Euler variables [2]-[6].

# 3. Riemann's invariants and simple waves

In the previous section derived the exact solution of the Euler equations (1.1) and (1.2) in the form of a simple wave, which is described analytically in the form of the explicit dependence (2.3)-(2.6).

Let us now consider the solution of the Euler equations in a more general form than the simple wave, when the motion of the medium can be seen in the assumption of isentropicity for the case of a polytropic environment. In this case, the Euler equations (1.1), (1.2) and the equation of state (1.3) are reduced to the following relatively simple equations for the known Riemann invariants [13] (see equations (104.4) in [13]):

$$\left[\frac{\partial}{\partial t} + (V + c)\frac{\partial}{\partial x}\right]J_+ = 0;$$
$$\left[\frac{\partial}{\partial t} + (V - c)\frac{\partial}{\partial x}\right]J_- = 0 \qquad (3.1)$$

$$J_+ = V + \frac{2}{\gamma - 1}c; J_- = V - \frac{2}{\gamma - 1}c \qquad (3.2)$$

At the same time, the representation (3.2) for the special case under consideration of a polytropic medium admits a generalization that has the well-known form [13] (see (104.2) in [13]):

$$J_+ = V + \int\frac{dp}{\rho c}; J_- = V - \int\frac{dp}{\rho c} \qquad (3.3)$$

In (3.1), the differential operators acting on the Riemann invariants (3.2) or (3.3) are differentiation operators in the plane $(x, t)$ along two characteristics on which the corresponding value of $J_+$ or $J_-$ from (3.2) or (3.3) remains constant [13]. From the comparison (3.3) with (1.4), it follows that the Riemann invariants (3.3) coincide with those values that are constant in simple waves along the entire region of motion of the medium during the entire time, and not only on the characteristics [13]. For example, for the case of a polytropic medium, we obtain from the comparison (3.2) with (1.12) that the Riemann invariants in the case of a simple wave have the form:

$$J_- = -\frac{2c_{V=0}}{\gamma - 1} = const; J_+ = \frac{2c_{V=0}}{\gamma - 1} = const \qquad (3.4)$$

Let us now consider in (3.1) and (3.2) a more general case than the motion of a medium in the form of simple nonlinear Riemann waves. In this case, let the polytrope index have a value $\gamma = 3$ that is used, as is known, for example, to describe the explosion products of condensed explosives [50], [51]. In this case, equations (3.1) exactly coincide with the Hopf equations (1.9), which thus describe the general case of isentropic motion of a polytropic medium with the specified polytropic index, and not just the evolution of a simple wave. In this case, the general solution of Equations (3.1) for $\gamma = 3$ has the form [2]-[6]:

$$J_{\pm}(x,t) = \int_{-\infty}^{\infty} d\xi J_{0\pm}(\xi) \left(1 + t \frac{dJ_{0\pm}}{d\xi}\right) \delta(\xi - x + tJ_{0\pm}(\xi));$$
$$J_{\pm} = V(x,t) \pm c(x,t); J_{0\pm} = V_0(x) \pm c_0(x) \tag{3.5}$$

The exact solution (3.5) is a description of the Riemann invariants, which gives them an explicit analytical dependence on arbitrary initial conditions for the velocity field of the medium and the local speed of sound, which are set independently of each other, in contrast to the case of a simple wave.

In this case, the corresponding explicit analytical representations for the velocity of the medium and the local speed of sound are obtained from (3.5)

$$V(x,t) = \frac{1}{2}\left(J_{+} + J_{-}\right); c(x,t) = \frac{1}{2}\left(J_{+} - J_{-}\right); \rho(x,t) = \rho_{00}\left(\frac{c(x,t)}{c_{00}}\right); p(x,t) = p_{00}\left(\frac{c(x,t)}{c_{00}}\right)^{3}. \tag{3.6}$$

The obtained solution can be used to develop examples of the application of Riemann invariants known in hydrodynamics and aerodynamics for the Riemann solvers [16] and others applications [50]. In particular, in [50], the Riemann invariants are used as new variables in the problem of bubble collapse in a compressible fluid, for which the equation of state corresponds to the polytropic exponent $\gamma = 3$ when the value of the local speed of sound is proportional to the density of the fluid $c \propto \rho$ as in (3.6).

## 4. The time of shock wave arising.

The exact solutions obtained for hydrodynamic fields in the form (2.3) - (2.6) and (3.5), are defined only for a finite time interval, up to the time of their collapse. Actually, the solutions lose smoothness at some time instance $t = t_0$, depending on the initial velocity field distribution. The value $t_0$ is defined as the minimum time such that the following equality holds (see also [4-6]):

$$1 + t(\frac{dV_0(x)}{dx} \pm \frac{dc_0(x)}{dx}) = 0;$$

$$t_0 = \frac{1}{\max\limits_{x=x_M}\left|\dfrac{du_0(x)}{dx}\right|}; u_0 \equiv V_0 \pm c_0 \qquad (4.1)$$

The initial field of the local speed of sound in (4.1) for the solution (3.5) is independent of the initial speed field. Therefore, the collapse time of this exact solution of the Euler equations may depend on the choice of the sign in (4.1). To solve the Euler equations in the form of a simple wave (2.3)-(2.6), the initial condition for the local speed of sound already depends on the initial velocity field. In this case, for example, for the case of a polytropic medium with an isentropic equation of state (1.3), this dependence has the form (2.16), from which it follows that the collapse time (4.1) is independent of the direction of propagation of a simple wave.

Further, let us consider for simplicity only the exact solution for a simple wave (2.3)-(2.6). For the case of a polytropic medium, taking into account (2.17), we obtain that the minimum time after which the collapse occurs for a simple wave has the following dependence on an arbitrary initial velocity field:

$$t_0 = \frac{2}{(\gamma+1)\left(\max\limits_{x=x_M}\left|\dfrac{dV_0}{dx}\right|\right)} \qquad (4.2)$$

In (4.2), the minimum time $t_0$ during which the nonlinear collapse of a simple wave occurs at the point with the coordinate $x = x_M$ where the function $|dV_0/dx|$ reaches its maximum value is determined

Thus, solution (2.3) - (2.6) is defined and conserves smoothness only on the time interval

$$0 \le t < t_0 \qquad (4.3)$$

That is why, for example, in (2.14), the integral on the right-hand side converges only under the condition (4.3). Thus, for arbitrary initial velocity fields, an explicit analytical representation for the collapse time of a nonlinear simple wave is established, which is described as an explicit exact analytical solution of the 1-D Euler equation (2.3)-(2.6) or (3.5). Note that in [13] an implicit Riemann solution is used to determine the collapse time, which does not allow obtaining the sort of explicit analytical dependence for arbitrary initial velocity fields such as the one presented in (4.2). Only in the cases of initial conditions when from the function $V_0(x)$ it is possible to obtain the analytical representation for the reverse function $x = f(V_0)$ can this be done on the basis of the theory [13]. Indeed, let us consider the following two examples of initial conditions, when only in the first example the method developed in [13] allows us to give an explicit expression for the collapse time that coincides with the value obtained from Equation (4.2).

For example, from (4.2) for the initial velocity distributions $V_0(x) = a \exp(-x^2 / 2x_0^2)$, the minimal time of the shock wave arising is (when $x_M = x_0 > 0$ in (4.2)):

$$t_0 = \frac{2x_0 \sqrt{e}}{a(\gamma + 1)} \qquad (4.4)$$

The value of the shock wave arising time (4.4) is exactly the same as may be obtained from the Landau theory [13] (from formula (101.13) in [13]).

Let us consider another case when the initial velocity field takes the asymmetrical form $V_0(x) = a_1 \frac{x}{x_1} \exp(-x^2 / 2x_1^2)$ for which the representation $V_0(x) = -V_0(-x)$ is valid.

In this case, from (4.2) we obtain that the collapse of a simple wave occurs at points with coordinates $x_M = \pm x_1 \sqrt{3}$ and at the same time the minimum time of collapse implementation has the form:

$$t_0 = \frac{x_1 e^{3/2}}{a_1(\gamma + 1)} \qquad (4.5)$$

On the basis of the implicit Riemann solution and on the corresponding Landau theory [13] it is impossible to obtain this explicit analytical representation (4.5) for the time of the shock wave arising also for that asymmetrical case of an initial velocity field.

## 5. Solution regularization by dissipation and shocks

### Threshold homogeneous dissipation

It is well known [13] that the presence of viscosity and thermal conductivity can lead to the dissipation of the wave energy. Here we shall take into account the dissipation of simple wave just as it is considered in [13] on the example of sound absorption due to these factors (see formulas (79.6) and (81.10) in [13]). For simplicity, the dissipation is introduced by addition to the right side of the equations (1.5)-(1.9) of a term that provides an exponential decrease in time of the corresponding hydrodynamic fields like $\exp(-t\mu)$. Here the increment $\mu = const$ characterizes the value of dissipative factors in the absence of dispersion (see (81.10) in [13]).

In this case, for example, Equation (1.9) has the form:

$$\frac{\partial \widetilde{u}}{\partial t} + \widetilde{u}\frac{\partial \widetilde{u}}{\partial x} = -\widetilde{u}\mu \qquad (5.1)$$

In all other equations (1.5)-(1.8) it is also necessary to enter in the right hand side the linear term for the corresponding field.

Let us use the representation $\widetilde{u} = u\exp(-t\mu); \widetilde{\rho} = \rho\exp(-t\mu); \widetilde{p} = p\exp(-t\mu); \widetilde{c} = c\exp(-t\mu); \widetilde{V} = V\exp(-t\mu)$ in (5.1). This gives modified equations (1.5)-(1.8) for the fields $u; \rho; p; c; V$ which exactly match the original equations (1.5)-(1.9) if in (1.5)-(1.9) one replaces the time variable $t$ by a new variable $\tau(t)$ in the form:

$$t \to \tau(t) = \frac{(1 - \exp(-t\mu))}{\mu} \qquad (5.2)$$

Thus, if in all the solutions (2.3) - (2.6) (see also (2.17)) one replaces $t \to \tau(t)$, then their regularization in accordance with condition (4.3) may be provided only under the following generalization of the condition (4.3):

$$0 \le t < \widetilde{t}_0(\mu) = \frac{1}{\mu}\ln\left(\frac{1}{1 - t_0\mu}\right), 0 \le \mu < \mu_{th} = \frac{1}{t_0}; \widetilde{t}_0(0) = t_0 \qquad (5.3)$$

$$0 \le t < \infty, \mu \ge \mu_{th} = \frac{1}{t_0}$$

For arbitrary large time intervals, the regularization condition (5.3) is always satisfied for sufficiently large super threshold values of the dissipation coefficient:

$$\mu \geq \mu_{th} = \frac{1}{t_0} \qquad (5.4)$$

In (5.4), the value $t_0$ depends only on the initial conditions and is defined in (4.1) and in (4.2) for the case of a polytropic medium. Two examples for the application of (4.2) to different initial velocity are given in (4.4) and (4.5).

Thus, for the super threshold value (5.4) of the dissipative factor $\mu$, regularization of solutions (2.3)-(2.6) is possible.

To obtain an estimate of the value of the dissipation coefficient $\mu$, we consider the law of dispersion of sound waves in a linear approximation, taking into account the effects of viscosity and thermal conductivity, when the motion in the sound wave is no longer adiabatic [13]. In [13], the case of a medium with high thermal conductivity is considered, when the effect of viscous dissipation can be neglected in the limit of small Prandtl numbers (see Problem 3 in Paragraph 79 in [13]).

The generalization for this solution of the problem [13], in which we additionally take into account the effects of viscosity at a finite value of the Prandtl number, is obtained. In this case, for the 1-D case, in the linear approximation for the entropy, temperature, density, pressure, and velocity perturbation fields $s'; T'; \rho'; p'; V'$ from the entropy balance, continuity, and Navier-Stokes equations for a compressible medium, we have the following system of equations:

$$\frac{\partial s'}{\partial t} = \frac{c_p \chi}{T} \frac{\partial^2 T}{\partial x^2}; \frac{\partial \rho'}{\partial t} + \rho_0 \frac{\partial V'}{\partial x} = 0; \frac{\partial V}{\partial t} = -\frac{1}{\rho_0} \frac{\partial p}{\partial x} + \left( \frac{4}{3} \nu_1 + \nu_2 \right) \frac{\partial^2 V}{\partial x^2} \qquad (5.5)$$

In (5.5) $\chi$ is the coefficient of thermal conductivity, $c_p$ is the heat capacity at constant pressure, $\nu_1$ and $\nu_2$ are the coefficients of kinematic shear and volumetric (second) viscosity, respectively. As in [13], for the closure of the system (5.5), we use the following thermodynamic relations:

$$\rho' = \left(\frac{\partial \rho}{\partial T}\right)_p T' + \left(\frac{\partial \rho}{\partial p}\right)_T p'; s' = \left(\frac{\partial s}{\partial T}\right)_p + \left(\frac{\partial s}{\partial p}\right)_T p' \tag{5.6}$$

We will look for the solution of the system (5.5), (5.6) in the form of an expression proportional to $\exp(i(kx - \omega t))$.

In this case, we use the following known thermodynamic relations [52] (see relations (13.6), (16.4) and (16.9) in [52]):

$$\left(\frac{\partial \rho}{\partial p}\right)_T = \frac{1}{c_T^2}; c_T^2 = c^2/\gamma; \gamma = c_p/c_v; c^2 = \left(\frac{\partial p}{\partial \rho}\right)_s;$$
$$\left(\frac{\partial s}{\partial T}\right)_p = \frac{c_p}{T}; \left(\frac{\partial s}{\partial p}\right)_T = -\left(\frac{\partial(v)}{\partial T}\right)_p = -\left(\frac{c_p - c_v}{TmN}\right)^{1/2}\frac{v}{c_T}; \left(\frac{\partial \rho}{\partial T}\right)_p = -\left(\frac{c_p - c_v}{TmN}\right)^{1/2}\frac{\rho}{c_T} \tag{5.7}$$

In (5.7), $v = mN/\rho$ is the volume occupied by the $N$ particles of the medium if the mass of the particle is equal to $m$. In this case, from (5.5)-(5.7) we obtain the following dispersion equation:

$$k^4\left(1 - \frac{i\omega\chi \Pr}{c_T^2}\right) - k^2\omega^2 c_T^{-2}\left(1 + ic_T^2/\omega\chi - i\Pr/\gamma\right) + i\omega^3/c_T^2\gamma\chi = 0 \tag{5.8}$$

In (5.8), $\Pr = \left(\frac{4}{3}\nu_1 + \nu_2\right)/\chi$ is Prandtl's number. In the limit $\Pr \to 0$, the dispersion equation (5.8) exactly coincides with the dispersion equation given in [13] (see Equation (3) in Problem 3 of Paragraph 79 in [13]). In this limit $\Pr \to 0$, in [13] for the case of high frequencies $\omega >> c_T^2/\chi$, the following two solutions of the dispersion equation (5.8) are obtained, which, when considering the real values of the wave number $k$, have the form [13] (see page 429 in [13] and the footnote on that page):

$$\omega = \omega_1 = -i\mu; \mu = \mu_1 = \chi k^2\gamma \tag{5.9}$$

$$\omega = \omega_2 = c_T k - i\mu; \mu = \mu_2 = \frac{c_T^2(\gamma - 1)}{2\chi\gamma} \tag{5.10}$$

From (5.8), for arbitrary values of the Prandtl number in the long-wave limit $k << \omega/c_T$, up to terms of order $O(k^3); k \to 0$, we obtain:

$$\omega = \omega_3 = \pm kc_T\sqrt{\gamma} - k^2\chi \Pr/2 - i\mu_3 + O(k^3); \mu = \mu_3 = \chi k^2\gamma/2 \tag{5.11}$$

In (5.11) there is no restriction on the value of the Prandtl number, which distinguishes this solution from the solution presented in [13] and having the form (5.9) or (5.10). However, in the under consideration limit of long waves, the exponential attenuation of small perturbations in (5.11) and in (5.9) differ only twice quantitatively and qualitatively have the same physical meaning

From condition (5.4), taking into account (5.10) for the case of the initial velocity field corresponding to the minimum time of discontinuity occurrence (4.4), we obtain the following relation between the thermal Reynolds and Mach numbers, in which the solution preserves smoothness over an unlimited time interval:

$$\mathrm{Re}_\chi > \mathrm{Re}_{1th} = \frac{M_T^2}{\sqrt{e}}\left(\frac{\gamma+1}{\gamma-1}\right) \tag{5.12}$$

From the other side, the solution collapses for the sub-threshold value the thermal Reynolds number when $\mathrm{Re}_\chi < \mathrm{Re}_{\chi th}$. In (5.12) $\mathrm{Re}_\chi = \frac{x_0 a}{\chi}; M_T = \frac{a}{c_T}$ is the definition of the thermal Reynolds and Mach numbers, respectively. For the case of an asymmetric initial velocity field, for which the collapse time is defined in (4.5), the condition (5.12) is supplemented only by the factor $2/e$ on the right side (5.12), if $a_1 = a; x_1 = x_0$.

The conclusion about the possibility of the existence of the smooth solution on the infinite time interval for the case under consideration of a one-dimensional simple wave is unexpected even in the case of finite dissipation. Indeed, previously it was assumed a priori the absence of a positive decision of the problem on existence of solutions of Euler's equation and Navier-Stokes for the case of a compressible medium in all cases [11].

Note that instead of the condition for preserving the smoothness of the solution in the form (5.12) for the case of solution (5.9), we already get a restriction from above on the value of the thermal analog of the Reynolds number:

$$\mathrm{Re}_\chi < \mathrm{Re}_{2th} = \frac{2k^2 x_0^2 \gamma \sqrt{e}}{\gamma+1} \tag{5.13}$$

In the case of the solution (5.11), for arbitrary Prandtl numbers (although only in the long-wave limit), the smoothness condition of the solution over an unlimited time interval has the form similar to (5.13), but with a threshold Reynolds number half the size of the one in (5.13).

## Shock wave velocity

The integral in (2.14) that diverges when the shock wave occurs when condition (5.4) is violated becomes finite when condition (5.4) is fulfilled. Indeed, a shock wave is characterized by a jump in density at its front, and the value (2.14) can determine the measure of smoothness of the solution before the shock wave occurs and at the moment of its occurrence $t = \tilde{t}_0(\mu)$ according (5.3). Taking into account the dissipation and the corresponding regularization of the solution, the integral (2.14) allows us to determine the evolution of the smoothed gap zone. As noted in [13], the shock wave front in reality has a finite width associated with the effects of viscosity and thermal conductivity. When such dissipative factors are taken into account, the expression (2.14) has the following representation:

$$(\frac{\partial \tilde{\rho}}{\partial x})^2 = e^{-2\mu t} \int_{-\infty}^{\infty} d\xi \frac{(d\rho_0 / d\xi)^2 \delta(\xi - x + \tau(t)u_0(\xi))}{1 + \tau(t)du_0 / d\xi};$$
$$\tau(t) = \frac{1 - \exp(-t\mu)}{\mu}$$

(5.14)

Under the condition (5.4), the integral (5.14) remains finite for any moment of time.

Let us consider the condition $0 < (\mu - \mu_{th}) / \mu_{th} = \varepsilon \ll 1$ when the main contribution to the integral (5.14) is made within the limits of integration in which the denominator is arbitrarily small but not zero, in accordance with the condition (5.4). With this in mind, using the example of specific symmetrical initial conditions corresponding to the estimates (4.4) for the value $t_0$, we obtain an estimate of the evolution in time of the integral (5.14) for any moment in time. Note that the symmetric initial velocity distribution used to obtain the estimate (4.4) corresponds to the nature of the motion of the medium observed in the underwater electric explosion of a thin wire in the experiment [53], the data of which are considered further for estimating the velocity of the shock wave, based on the relation (5.14).

In particular, for the initial conditions (2.15) and (2.16) corresponding to (4.4) from (5.14), one can get the estimation:

$$\left(\frac{\partial \widetilde{\rho}}{\partial x}\right)^2 \cong A e^{-2t\mu} \frac{\rho_\infty^2}{x_0^2} \delta(\frac{x - x_{sh}(t)}{x_0})$$

$$x_{sh}(t)/x_0 = 1 + \frac{\tau(t)}{t_0}(1 \pm \frac{c_\infty t_0}{x_0}); \qquad\qquad (5.15)$$

$$A = \frac{a^2}{ec_\infty^2} \frac{\left(1 \pm \frac{(\gamma-1)a}{2c_\infty \sqrt{e}}\right)^{\frac{3-\gamma}{\gamma-1}}}{(1 - \tau(t)/t_0)}$$

According to (5.8), the velocity of propagation of a regularized discontinuity has representation:

$$D = \frac{dx_{sh}}{dt} = e^{-t\mu}\left(c_\infty + \frac{x_0}{t_0}\right) \qquad\qquad (5.16)$$

Let us consider in (5.16) the value $\mu \approx 0.24 \times 10^6 \sec^{-1}$ which can be taken from data [53], where the velocity of the shock wave decreases from 3000 m/sec to 2100 m/sec during $1.5 \times 10^{-6}$ sec for the case of the underwater electrical explosion of a single wire.

For example, for the case of an explosion of a copper wire with an initial radius $x_0 = 0.3 \times 10^{-3}$ m considered in [53], an estimate of the time of occurrence of the shock wave $t_0 \approx 0.15 \times 10^{-6}$ sec can be obtained based on the formula (4.4) and near the same value $\widetilde{t}_0(\mu) \approx 0.153 \times 10^{-6}$ sec from (5.3). For this purpose, the formulas (4.4) and (5.3) uses the maximum velocity of the medium observed in [39] behind the shock wave front, estimated by the value $a \approx 800$ m/sec, as well as the adiabatic index $\gamma = 7.15$ for water [14], [53].

Taking this estimate $t_{00}$ into account, we put $t\mu \cong t_0\mu \approx 0.04$ in (5.16). In this case, as well as for $c_\infty = 1483$ m/sec [54], we obtain an estimate for the shock wave velocity $D \approx 3325$ m/sec from (5.16). It corresponds, with an accuracy of up to 10% , to the maximum measured value of the shock wave velocity observed in [53] and given above. Note also that the above estimate obtained on the basis of (5.16) corresponds with approximately the same accuracy to the data given in [54] (see Table 3 in [54]), where the velocity of the medium behind the shock wave front $a = 812$ m/sec corresponds to the velocity of the shock wave $D = 3083$ m/sec.

## Stochastic modeling of effective viscosity

The dissipation mechanism discussed above is homogeneous, for which the possibility of threshold regularization of the solution of the Hopf equation was also established in [4]-[6]. At the same time, when modeling the effects of molecular viscosity in [4]-[6], the possibility of regularization of the solution is shown without the presence of any threshold in terms of the kinematic viscosity coefficient $\nu > 0$. This conclusion also applies to the regularization of the exact solution of the Euler equations under consideration. In this case, for example, the result of such regularization for the velocity field (2.17) leads to the following kind of exact solution, which preserves smoothness over an unlimited time interval for any non-zero viscosity coefficient:

$$\langle V(x;t) \rangle = \int_{-\infty}^{\infty} dB \frac{\exp(-B^2/4t\nu)}{2\sqrt{\pi t \nu}} V(x-B,t) \qquad (5.17)$$

In (5.17), we present a statistical averaging of the solution for the velocity field $V(x-B;t)$ from (2.17), where

$B(t) = \int_0^t dt_1 \tilde{V}(t_1); \langle \tilde{V} \rangle = 0; \langle \tilde{V}(t)\tilde{V}(t_1) \rangle = 2\nu\delta(t-t_1)$. The modeling of the viscosity, as in [4]-[6], carried out by replacing $u\dfrac{\partial u}{\partial x} \to (u + \tilde{V}(t))\dfrac{\partial u}{\partial x}$ in the equation (1.9). Thus the transfer by a random Gaussian velocity field $\tilde{V}(t)$, for which the formula for Furutsu-Novikov $\left\langle \tilde{V}(t)\dfrac{\partial u}{\partial x} \right\rangle = -\nu\dfrac{\partial^2 u}{\partial x^2}$ is valid, is introduced (see references in [5], [6]).

The specified modifications to equation (1.9) has an exact solution in the form (2.17), where must only be replaced $x \to x - B$. After that, the random field $B$ is averaged to obtain the solution (5.17).

## 6. Turbulence theory

### Structure functions and dissipative anomaly

In the theory of turbulence, one of the main characteristics of a turbulent flow is the structure function of the hydrodynamic field under study [9]. It allows an average description of a complex turbulence phenomenon that manifests itself in random pulsations of the medium flow. Until now, as noted in the Introduction, it is generally accepted that it is still far from complete to create a theory that allows one to calculate all possible correlations of such pulsations from the first principles - that is, from solutions of the hydrodynamic equations. According to experimental data and numerical simulation [55]-[57], the deviation from the Kolmogorov turbulence theory estimations increases with increasing order of the structure function. The reason for this difference, which is related to fluctuations in the rate of energy dissipation, is the subject of modern research on the phenomenon of intermittency of turbulent flows [7], [9], [13], [55]-[57].

Below it is shown that any single-point and multi-point correlation functions and corresponding spectra characterizing the turbulence regime can be obtained directly from the exact explicit solution (2.17) of one-dimensional Euler equations.

Based on the exact solution (2.17), we consider the structure function of the velocity field, which we define as:

$$S_p(r;t) = \frac{1}{L}\int\limits_{-\infty}^{\infty}dx\big[V(x+r;t)-V(x;t)\big]^p \qquad (6.1)$$

In (6.1) $p = 2;3;4..$ and a characteristic integral length scale $L = R_1^{-1}(0)\int\limits_{-\infty}^{\infty}drR_1(r;t); R_1 = \int\limits_{-\infty}^{\infty}dxV(x+r;t)V(x;t)$ [9]

(see (12.21) in vol.2 [9]) is used.

Without taking into account the dissipation, the solution of the Euler equation (2.17) has the following infinite set of motion invariants $\int\limits_{-\infty}^{\infty}dxV^m(x;t) = \int\limits_{-\infty}^{\infty}dxV_0^m(x); m = 1,2,...$

Thus, for the solution (2.17) integral length scale has also invariant representation:

$$L = \frac{P_0^2}{E_0}; P_0 = \int\limits_{-\infty}^{\infty}dx|V_0(x)|; E_0 = \int\limits_{-\infty}^{\infty}dxV_0^2(x) < \infty; V_0(x) = V(x;t=0) \qquad (6.2)$$

For example, for the initial symmetrical distribution of the velocity field $V_0(x) = a\exp\left(-x^2/2x_0^2\right)$ from (6.2), it follows that the characteristic length scale $L = L_0 = 2\sqrt{\pi}x_0$. For the case of an asymmetric distribution of the initial velocity field $V_0(x) = a_1 \dfrac{x}{x_1} \exp(-\dfrac{x^2}{2x_1^2})$, we obtain $L = L_1 = 8x_1/\sqrt{\pi}$ from (6.2). In particular, for the same value

$x_1 = x_0$, we have $L_0/L_1 = \pi/4 < 1$.

**Dissipative anomaly in the compressible turbulence**

  Before considering the exact solutions for the two-point structural functions (6.1), we give the exact solutions for the one-point moments of the velocity field and the spatial derivatives of this field, which also gives an exact solution to the problem of the dissipative anomaly for turbulence in a compressible medium.

  Note that near the collapse of the solution in the limit $t \to t_0$, the integrals $\dfrac{1}{L}\displaystyle\int_{-\infty}^{\infty} dx V^m(x,t)$ are also time-invariant.

At the same time, local in a small neighborhood $[x_M - \beta; x_M + \beta]$, $\beta << x_M$ near the singularity in the limit

$t_\nu = t_0 - t \to 0$ an increase is realized for the integral of the square of the velocity in the form:

$$\frac{d}{dt}\left(\frac{1}{L}\int_{x_M-\beta}^{x_M+\beta} dx V^2(x;t)\right) \cong 4\beta \frac{V_0^2(x_M)}{t_0 L} + o(\beta/x_M) \qquad (6.3)$$

  Representation (6.3) is obtained according to solution (2.17) and by taking into account the identities (2.7) and

(2.8). In this case, it is also necessary to take into account the relations used to obtain an estimate of the minimum

collapse time of the solution in (4.2):

$$\left(dV_0/dx\right)_{x=x_M} = -2/t_0(\gamma+1); \left(d^2V_0/dx^2\right)_{x=x_M} = 0; \left(d^3V_0/dx^3\right)_{x=x_m} \equiv V_0''' > 0; \left(dV_0/dx\right)_{x=x_M} \equiv V_0' < 0. \quad (6.4)$$

This increase in the integral in (6.3) is exactly compensated by its decrease on intervals $\left(-\infty; x_M - \beta\right]$ and

$\left[x_M + \beta; \infty\right)$.

  Let us consider the integral kinetic energy of the turbulent flow of a compressible medium

$E_C = \dfrac{1}{2L}\displaystyle\int_{-\infty}^{\infty} dx \rho(x;t) V^2(x;t)$. In the general case, at a finite viscosity, the equation of the balance of the integral

kinetic energy has the form [5] (see (3.5) in [5]):

$$\varepsilon_C \equiv \frac{1}{\rho_\infty}\frac{dE_C}{dt} = -I_D + I_P \; ;$$

$$I_D = \nu_D \Omega_2 = \frac{\nu_D}{L}\int\limits_{-\infty}^{\infty}dx\left(\frac{\partial V(x;t)}{\partial x}\right)^2 ; I_P = \frac{1}{\rho_\infty L}\int\limits_{-\infty}^{\infty}dx\, p(x;t)\frac{\partial V(x;t)}{\partial x}; \nu_D = (\frac{4\eta}{3}+\varsigma)/\rho_\infty \tag{6.5}$$

In (6.5) $\Omega_2$- is the enstrophy, $\eta$-and $\varsigma$ are the shear and volume (second) viscosity coefficients, respectively.

At zero viscosity, the balance equation (6.5) coincides with the equation given in [58] (see (4) in Chapter 10 [58]). In this case, for the exact solution of the Euler equations (2.17) and (2.18) in the region of regularity of the solution $0 \le t < t_0$ and in the limit $t \to t_0$ the kinetic energy is invariant in time, because

$$I_P = \frac{p_\infty}{L}\int\limits_{-\infty}^{\infty}dx\left(1 \pm \frac{(\gamma-1)}{2c_\infty}V_0(x)\right)^{\frac{2\gamma}{\gamma-1}}\frac{dV_0}{dx} = 0\text{ for any initial velocity field satisfying zero boundary conditions at}$$

infinity, according to (6.2). Here the representation (2.16) is used for the initial distributions of the pressure field through the initial distribution of the velocity field.

Let us consider only the first term of the balance equation (6.5) in the limit of arbitrarily small viscosity. As a first step, from (2.17) an exact solution for the enstrophy generalization of any order $n$ is obtained:

$$\Omega_n = \frac{1}{L}\int\limits_{-\infty}^{\infty}dx\left(\frac{\partial V(x,t)}{\partial x}\right)^n = \frac{1}{L}\int\limits_{-\infty}^{\infty}dx\left(\frac{dV_0(x)}{dx}\right)^n\left(1+t\,\frac{(\gamma+1)}{2}\frac{dV_0(x)}{dx}\right)^{-(n-1)} ; n = 2,3,4,... \tag{6.6}$$

In the limit $t \to t_0$, the main contribution to the integral (6.6) will be given by a narrow region $-\beta \le x - x_M \le \beta$ near the point $x = x_M$, which is corresponded to the divergence of the integral (6.6) in the limit $t_\nu = t_0 - t \to 0$.

Let us consider the first three terms (see (6.4)) in the decomposition of the function $dV_0/dx$ in the denominator (6.6) into a Taylor series near a point $x = x_M$. Thus, the generalization of ensrtophy in (6.6) in this limit takes form:

$$\Omega_n = 2\pi\frac{(V_0')^n l_0}{L}\left(\frac{t_0}{t_0-t}\right)^{n-\frac{3}{2}}arctg\left(\frac{\beta}{l_0}\sqrt{\frac{t_0}{t_0-t}}\right);$$

$$l_0^{-2} = (n-1)V_0'''/2|V_0'| \tag{6.7}$$

For example, in (6.7) there is a singularity $\Omega_n \propto O\left(1/(1-t/t_0)^{n-\frac{3}{2}}\right)$ in the limit $t \to t_0$ and $1 >> \dfrac{\beta}{l_0} >> \sqrt{\dfrac{t_0-t}{t_0}}$ where $t_0$ is determined in (4.2). In this limit, there is also a relation $\Omega_{2n}/\Omega_n^2 \propto 1/(1-t/t_0)^{3/2} >> 1$ that is typical for the regime of strong intermittency of turbulence [7], [9], [10] and [55]-[57].

In this case the enstrophy $\Omega_2$ has singularity of the type $\Omega_2 \propto O\left(1/\sqrt{t_0-t}\right)$. Note that this dependence is in accordance with the following representation for the enstrophy singularity type $\Omega_2 \propto O\left(1/(t_0-t)^q\right); 1 \geq q \geq 1/2$ [59] (see (3.13) in [59]) obtained from the heuristic model of the turbulent energy cascade. For the given example of the exact solution following from (6.7), the lower limit value $q = 1/2$ in the condition given in [59] corresponds. On the other hand, in the limit $1 >> \sqrt{\dfrac{t_0}{t_0-t}} >> \dfrac{\beta}{l_0}$ it follows from (6.7) that $\Omega_2 \propto \dfrac{t_0}{t_0-t}$ and this corresponds to the upper limit value $q = 1$ in the theory [59].

From the case $n = 2$ in (6.7) for the first term in the integral kinetic energy balance equation (6.5), we obtain an asymptotic representation near the singularity in the limit of arbitrarily small viscosity:

$$\varepsilon_C = \lim_{\substack{\nu_D \to 0 \\ t \to t_0}} \frac{\nu_D \pi \left(V_0'(x_M)\right)^2 l_0}{L} \left(\frac{t_0}{t_0-t}\right)^{1/2} \to \varepsilon_{C0} = const > 0; \qquad (6.8)$$

In (6.8), the limit value will be finite if $\nu_D \left(\dfrac{t_0}{t_0-t}\right)^{1/2} \propto O(1)$. When taking into account the homogeneous dissipation, it is necessary to use the replacement (5.2) in (6.8). At the same time, it is necessary to consider both the limit $\nu_D \to 0$ and the limit $\mu \to 1/t_0; \mu > \mu_{th} = 1/t_0$. In this case, the finite value in the limit $\varepsilon_C \to \varepsilon_{C0}$ takes place in the case when $\nu_D (t_0 \mu - 1)^{-1/2} \propto O(1)$.

It follows from (6.5) and (6.8) that near the singularity in the limit of arbitrarily small viscosity, the rate of change of the integral kinetic energy $\varepsilon_{C0} = const$, which characterizes the dissipative anomaly of turbulence in a compressible medium.

Thus, based on the exact solution of the Euler equations, a condition is established under which the turbulence energy is dissipated near the collapse of the solution, even in the limit of zero viscosity. This gives an example of an explicit solution to the well-known problem of Onsager's dissipative anomaly in the turbulence theory [1], [7], [17]. Indeed, until now, this problem has been considered in the framework of various model approaches, which took into account only some properties of the hydrodynamic equations, but not the exact solutions of the Euler or Navier-Stokes equations themselves [60].

Note, however, that the dissipative anomaly occurs only in the presence of an irreducible singularity of the solution and is absent for the regular exact solution (5.17), where regularization is achieved by taking into account any arbitrarily small effective viscosity modeled by a random Gaussian delta - time-correlated velocity field.

In this case we obtain $\langle \varepsilon_C \rangle = \lim_{\nu \to 0} \nu \langle \Omega_2 \rangle \propto O\left(\nu^{5/6}\right) \to 0$; $\langle \Omega_2 \rangle = \int_{-\infty}^{\infty} dx \left( \frac{\partial \langle V(x;t) \rangle}{\partial x} \right)^2$ . In order to make sure of this, you need to use the view $\omega(k) = \int_{-\infty}^{\infty} dx e^{ikx} \frac{\partial \langle V \rangle}{\partial x} = e^{-k^2 \nu t \pm ikc_\infty} \int_{-\infty}^{\infty} dx \frac{dV_0}{dx} \exp\left(ikS(x;t)\right); S(x;t) = x + t \frac{(\gamma + 1)}{2} V_0(x)$ , where

$\langle \Omega_2 \rangle = \frac{1}{2\pi} \int_{-\infty}^{\infty} dk \omega(k) \omega(-k)$ . In the limit $\nu \to 0; t \to t_0$ this gives $\omega(k) \propto e^{-k^2 \nu t_0 \pm it_0 c_\infty} k^{-1/3}$ . Here we used substitution

$y = k \sqrt{t_0 \nu}$ in integral $\langle \Omega_2 \rangle$ and consideration of fast oscillation of the periodic function $\exp\left(i \lambda y S(x;t_0)\right)$ in the integral $\omega(k(y))$ due to the large parameter $\lambda = (t_0 \nu)^{-1/2} \to \infty$ in the limit $\nu \to 0$ (see the same details of derivation below in the estimation of spectra (6.17) and (6.20)).

Thus, only for the solution of the Euler equations at zero viscosity or its modification, which takes into account the homogeneous dissipation, it is possible to realize the dissipative anomaly. For the multi-dimensional case it is interesting to check this conclusion, first of all in the frame of the exact solution to Hopf's equations [4]-[6], which is taken into account effective viscosity.

## Structure functions p=2, 4

Let us consider in more detail the case when in (6.1) the exponent of $p = 2$, which corresponds to a second-order structural function. In this case, from (6.1) and (2.17) we obtain a representation for the structure function $S_2$ in the explicit analytical form:

$$S_2 = \frac{2}{L} \sum_{n=1}^{\infty} \frac{r^{2n}(-1)^{n+1}}{(2n)!} \int_{-\infty}^{\infty} dx \left( \frac{d^n V_0}{dx^n} \right)^2 =$$

$$= \frac{r^2}{L} \left( \int_{-\infty}^{\infty} dx \left( \frac{dV_0}{dx} \right)^2 - \frac{r^2}{12} \int_{-\infty}^{\infty} dx \left( \frac{d^2 V_0}{dx^2} \right)^2 + O(r^4) \right) \qquad (6.9)$$

For example, when $V_0(x) = a \exp\left( -x^2 / 2x_0^2 \right)$ from (6.9) we get that $S_2 = S_{20} = r^2 A_0 - O(r^4 / x_0^4); A_0 = a^2 / 4x_0^2$.

For the initial velocity field $V_0 = a_1 \frac{x}{x_1} \exp\left( -x^2 / 2x_1^2 \right)$ from (6.9), we obtain

$S_2 = S_{21} = r^2 A_1 + O(r^4 / x_1^4); A_1 = 3\pi a_1^2 / r_1^2$. At the same time, in the case of equality $a = a_1; x_0 = x_1$ we obtain that $A_1 / A_0 \approx 1.178$ and there is only a slight difference in the corresponding coefficients, despite the qualitative difference between the symmetry of the initial fields in the two cases under consideration.

Note that the solution (6.9) for a second-order structural function, as well as for even-order structural functions in general, turns out to be time-independent in an explicit form. However, for any point in time $t > 0$, the representation (6.9) differs from the structural function (6.1) at the initial time $t = 0$:

$S_2(r;t = 0) = \frac{r^2}{L} \left( \int_{-\infty}^{\infty} dx \left( \frac{dV_0}{dx} \right)^2 + \frac{r^2}{24} \int_{-\infty}^{\infty} dx \left( \frac{d^2 V_0}{dx^2} \right)^2 + O(r^4) \right)$. The coincidence of $S_2(r;t = 0)$ with (6.9) holds only for the first terms of the power series expansion $r^2$.

The first terms of the expansion of the fourth-order structural function at the initial moment of time

$S_4(r;t = 0) \cong \frac{r^4}{L} \int_{-\infty}^{\infty} dx \left( \frac{dV_0}{dx} \right)^4$ and the structural function $S_4(r;t > 0) \cong \frac{12r^4}{L} \int_{-\infty}^{\infty} dx \left( \frac{dV_0}{dx} \right)^4$ no longer coincide and their relation $S_4(t > 0) / S_4(t = 0) = 12$ in the limit $r \to 0$.

Thus, kurtosis $K(t>0) = \dfrac{S_4}{S_2^2} - 3$ for our solution is equal to $K(t>0) = 12K(0) + 33$ and gives the strong deviation from the usual known relation corresponding to Gaussian processes when $K=0$. This is known property of the strong turbulence in the intermittence regime [9]. For example, $K(t>0) \approx 9.72$ and $K(0) \approx -1.94$ in the case of initial velocity $V_0(x) = a\exp(-x^2/2x_0^2)$.

Note that for all even-order structural functions, exact solutions can be obtained that, like (6.9), are time-independent and have an asymptotic representation $S_{2m}(r) \propto r^{2m} + o(r^{2m}); m = 1,2,\ldots$ The linear dependence of the exponent on the order of the structural function may indicate a monofractal turbulence regime described by such structure functions [57]. The same regime is also stated from the Cassini observational turbulence data in the magnetosheath of Saturn [45].

In this case, there is a scale invariance of the kurtosis value $K$, the value of which characterizes the deviation of the turbulent regime from the Gaussian random process and the corresponding degree of intermittency of the turbulence.

Note that the above relations for the structural functions $S_2$ and $S_4$ are valid only until the collapse time of the solution (2.17) $0 \le t < t_0$, defined in (4.2).

The scale invariance of the kurtosis magnitude $K$ can be violated at the time of collapse. Indeed, at the moment of collapse, the second-order structural function decreases, in the limit $r \to 0$ significantly slower than in (6.9), and has the form $S_2 \propto r^{5/3} + O(r^{7/3})$. For this case the Holder parameter $h = 5/6$ because $S_2 \propto r^{2h}$. This follows from the estimation of the energy spectrum (see below in (6.17)) and the well-known relationship between the energy spectrum and the second-order structural function [9]. In this case, the structural functions of a higher even order

$S_p$ will no longer be proportional to $r^{5p/6}$ with linear dependence of exponent on number $p = 4,6,8\ldots$

Thus, the scale invariance of the kurtosis value $K \ne 0$ is breaking at the moment of the solution collapse. In addition, the above example of a dissipative anomaly near the collapse of the solution corresponds to the value of the Holder parameter $h = 5/6 > 1/3$. As a result, it turns out that the implementation of the dissipative anomaly for the case of one-dimensional turbulence in a compressible medium occurs at a significantly smoother velocity field than is assumed in the Onsager theory [17], [32]. At the same time, however, there is a correspondence with the condition for the realization of the dissipative anomaly, obtained in [33].

### Structural functions p=3

Let us now consider the third-order structural function in (6.1). For the exponent value $p = 3$ in (6.1), from (2.17) we obtain the representation for a third-order structure function if $0 \le t < t_0$:

$$S_3 = \frac{3(\gamma+1)t}{2L} \sum_{n=1}^{\infty} (-1)^n \frac{r^{2n-1}}{(2n-1)!} \int_{-\infty}^{\infty} dx \left( \frac{d^n}{dx^n} V_0^2 \right)^2 \qquad (6.10)$$

From (6.10), we obtain an asymptotic representation:

$$S_3 = -6r \frac{(\gamma+1)t}{L} E_1 \left( 1 - \frac{r^2}{6r_0^2} + O(r^4/r_0^4) \right) \qquad (6.11)$$

$$r << r_0 = (E_2/E_1)^{1/2}; E_2 = \int_{-\infty}^{\infty} dx \left( V_0^2 \left( \frac{d^2 V_0}{dx^2} \right)^2 + \frac{1}{3} \left( \frac{dV_0}{dx} \right)^4 \right); E_1 = \int_{-\infty}^{\infty} dx V_0^2 \left( \frac{dV_0}{dx} \right)^2 .$$

The relation (6.10) is valid for any values of the time $0 \le t < t_0$ until the collapse of the solution. Moreover, for all structural functions of odd order, as for (6.10) and (6.11), there is always a linear dependence not only on the distance- $r$ $S_p(r) \propto -(r+o(r))t; p = 2n+1, n = 1,2,3...$, as in Kolmogorov's 4/5-law, but also and on the time.

At the moment of collapse $t = t_0$, instead of (6.11), the asymptotic representation takes place:

$$S_3 = -C_S V_0(x_M) L \left( \frac{dV_0}{dx} \right)_{x=x_M}^2 r + O(r^3) \qquad (6.12)$$

$$C_S = \frac{108}{\pi L^2} \Gamma \left( \frac{1}{3} \right) \left( (x_M + V_0(x_M)/V_0') \frac{V_0'}{V_0'''} \right)^{2/3}$$

The representation (6.12) for the third-order structural function is finite at identically zero viscosity, when an integral kinetic energy of turbulence is invariant.

In contrast to Kolmogorov's law 4/5, the form of the structural function in the obtained exact solution (6.12) does not violate the symmetry with respect to the change in the time sign $t \to -t$ because $V_0 \to -V_0$ in this case. Thus, the physical anomaly noted in [7] in connection with Kolmogorov's law 4/5 does not arise when obtaining a representation (6.12) for a third-order structural function $S_3$ based on exact solutions of the Euler equations. Therefore, we can assume that the artificial origin of this anomaly is due to the closure hypothesis used in the derivation of the 4/5 law from the Karman-Howarth equation. Indeed, in deriving this law, the condition $\frac{\partial S_2}{\partial t} << \langle \varepsilon \rangle$ is used [13]. In turn, this condition arising in the limit $t \to \infty$ is incompatible with a solution of Euler's equations that has a collapse on a finite time interval.

**Intermittence and dissipation fluctuations**

To describe the small-scale intermittency of turbulence the fluctuations of the energy dissipation rate of the energy turbulence $\varepsilon(x,t) = \frac{\nu}{2}\left(\frac{\partial V(x,t)}{\partial x}\right)^2$, which are characterized by a structural function $b_D(r) = \langle \varepsilon(x+r;t)\varepsilon(x;t) \rangle - \langle \varepsilon \rangle^2$ is used [9], [13]. Based on the exact solution of the Euler equations (2.17) and taking into account (6.6), as well as the identities (2.7) and (2.8), we obtain the following exact explicit analytical representation for the structural function of fluctuations in the kinetic energy dissipation rate:

$$b_D(r;t) = \frac{\nu^2}{4L}\int_{-\infty}^{\infty}d\xi_1\int_{-\infty}^{\infty}d\xi_2 \frac{(dV_0/d\xi_1)^2(dV_0/d\xi_2)^2}{S(\xi_1;t)S(\xi_2;t)}\delta\big(S(\xi_2;t)-S(\xi_1;t)-r\big)-\bar{\varepsilon}^2; \qquad (6.13)$$

$$S(x;t) = x + t\frac{(\gamma+1)}{2}V_0(x); \bar{\varepsilon}(t) = \frac{\nu}{2}\Omega_2$$

The structural function (6.13) corresponds to the representation for the spectrum (see below (6.20)) with an exponent of -2/3, which results in an estimate $b_D(r;t=t_0) \propto r^{-1/3}$ that is fair in the limit $r \to 0; t \to t_0$.

## Turbulence spectra

### Energy spectrum

Let us consider the energy spectrum. To determine it, we introduce a correlation function, which for the exact solution under consideration of the Euler equations (2.17) has the form:

$$R(r) = \frac{1}{L}\int\limits_{-\infty}^{\infty} dx V(x+r;t)V(x,t) = \frac{1}{L}\int\limits_{-\infty}^{\infty} d\xi_1 \int\limits_{-\infty}^{\infty} d\xi_2 V_0(\xi_1)A(\xi_1)V_0(\xi_2)A(\xi_2)B(\xi_1;\xi_2;r);$$

$$A(x;t) = 1 + \frac{(\gamma+1)t}{2}\frac{dV_0}{dx}; B = \delta(\xi_2 - \xi_1 - r + \frac{(\gamma+1)t}{2}(V_0(\xi_2) - V_0(\xi_1)))$$

(6.14)

Note that for the correlation function (6.14), an equality $R(r) = R(-r)$, is always satisfied, which indicates the isotropy of this exact solution of the Euler equations for a compressible medium in the form of a simple nonlinear wave when using the equation of state of the medium (1.3).

The Fourier transform of the correlation function (6.14) allows us to obtain the following exact closed representation for the energy spectrum, expressed in terms of an arbitrary initial velocity field in the form:

$$E(k) = \frac{1}{2\pi}\int\limits_{-\infty}^{\infty} dr R(r)\exp(-ikr) = \frac{1}{2\pi L}I(k)I^*(k)$$

(6.15)

$$I(k) = \int\limits_{-\infty}^{\infty} dx V_0(x)\frac{\partial S}{\partial x}\exp(ikS(x,t)) = \frac{i}{k}\int\limits_{-\infty}^{\infty} dx \frac{dV_0}{dx}\exp(ikS); I^*(k) \equiv I(-k)$$

$$S(x;t) = x + \frac{(\gamma+1)t}{2}V_0(x); A(x,t) \equiv \frac{\partial S}{\partial x}$$

In this case, the total energy per unit mass has the form $e = \int\limits_{0}^{\infty} dk E(k)$. A similar expression can be obtained for the density spectrum. In this case, we must replace the integral $I(k)$ in (6.15) with the integral

$$I_\rho(k) = \int\limits_{-\infty}^{\infty} dx \rho_0(x)A(x,t)\exp\left(ik\left(x + t\frac{(\gamma+1)}{2}V_0(x)\right)\right).$$ The initial distribution $\rho_0(x)$ in this integral is completely

determined by the type of initial velocity field according to the formula (1.12) and, in particular, from (2.15). Therefore, the dependence of the energy spectrum on the wave number obtained further determines the dependence on the wave number for the density pulsation spectrum and similarly for the pressure pulsation spectrum and the local sound velocity pulsation spectrum.

Let us consider the energy spectrum (6.15) in the limit $t \to t_0$ when the collapse of the solution (2.3) is realized at $A \to 0; \dfrac{\partial^2 S}{\partial x^2} \to 0; x \to x_M$, when taking into account the definition of the minimum collapse time $t_0$ given in (4.2).

At the same time, in the limit $k x_M \gg 1$ for estimating the integral $I(k)$ in (6.15), the stationary phase method [61] can be used, when the main contribution to it is made by the region near the value $x = x_M$ (see (4.2)). In this case, the representation of the function in the limit $t \to t_0$:

$$S(x) = S(x_M) + S'(x_M)(x - x_M) + \frac{1}{2!}S''(x_M)(x - x_M)^2 + \frac{1}{3!}S'''(x_M)(x - x_M)^3 + .. =$$

$$= (x - x_M)\left(1 - \frac{t}{t_0}\right) + \frac{(x - x_M)^3}{3!}t\frac{(\gamma+1)}{2}\left(\frac{d^3 V_0}{dx^3}\right)_{x=x_M} + O\left(\frac{(x - x_M)^4}{4!}t\frac{(\gamma+1)}{2}\left(\frac{d^4 V_0}{dx^4}\right)_{x=x_M}\right)$$ in (6.15) is used, when

taking into account the relations $S' = (\partial S / \partial x)_{x=x_M} = 0; S'' = (\partial^2 S / \partial x^2)_{x=x_M} = 0$ that follows from (4.2).

In this case, taking into account that the main contribution to the integral $I(k)$ in (6.15) gives a region near $x = x_M$, we obtain the following representation for this integral:

$$I(k) \cong \frac{i}{k}\left(\frac{dV_0}{dx}\right)_{x=x_M}\exp\left(ikS(x_M;t=t_0)\right)I_1(k)$$

$$I_1(k) = \int_{-\infty}^{\infty}dx\exp\left(ik\frac{S'''(x_M)}{3!}(x - x_M)^3\right) = \frac{2\sqrt{\pi}\Phi(0)}{k^{1/3}3^{1/3}}\left(S'''(x_M)/3!\right)^{-1/3} + O(k^{-2/3});$$

$$\Phi(x) = \frac{1}{\sqrt{\pi}}\int_{0}^{\infty}du\cos\left(ux + \frac{u^3}{3}\right); S'''(x_M) = V_0'''(x_M)/|V_0'(x_M)|; V_0'''(x_M) = \left(d^3 V_0/dx^3\right)_{x=x_M} > 0$$

(6.16)

Where after replacing the variables $y = \left(\dfrac{kS'''}{3!}\right)^{1/3}(x - x_M)$ in integral $I_1(k)$ it is expressed in terms of the Airy function with the argument equal to zero [62] (see (b.8) in [62]) and has an order $I(k) \propto O(k^{-4/3})$ for large wave numbers.

As a result, from (6.16) for the energy spectrum (6.15) with an arbitrary initial velocity field, we obtain a universal dependence on the wave number in the form of the law -8/3:

$$E(k) = C_E k^{-8/3} + O(k^{-10/3}) \qquad (6.17)$$

$$C_E = \frac{2^{1/3}}{L} \left( \frac{dV_0}{dx} \right)_{x=x_M}^{8/3} \left( \frac{d^3 V_0}{dx^3} \right)_{x=x_M}^{-2/3} \Phi^2(0)$$

In (6.17) $\Phi(z) = \sqrt{\pi} Ai(z)$ - the Airy function and [62] (see formula (b. 8) on page 784 in [62] where

$\Phi(0) = \dfrac{\sqrt{\pi}}{3^{2/3} \Gamma(2/3)} \approx 0.629$ ).

For example, in the case of initial velocity field $V_0 = a \exp(-x^2/2x_0^2)$ from (6.17) it is possible to obtain

$C_E = C_{E0} = \dfrac{a^2 \Phi^2(0)}{x_0^{5/3} e \sqrt{\pi}}$. Under another initial condition $V_0 = a_1 \dfrac{x}{x_1} \exp(-x^2/2x_1^2)$, which is asymmetric with respect to the origin, we obtain $C_E = C_{E1} = \dfrac{a_1^2 \Phi^2(0)}{x_1^{5/3} 3^{2/3} e^3}$ in (6.17).

Thus, for example, under the condition $a_1 = a; x_1 = x_0$ we get that $C_{E1}/C_{E0} = e^{-2} 3^{-2/3} \sqrt{\pi} \approx 0.115$. At the same time, regardless of the initial velocity field, the exponent in (6.17) has the same universal value of 8/3.

Universality of the exponent -8/3 in (6.17) is stated here not only on the absence of dependence of its magnitude from the initial velocity field, but also on the absence of the dependence from the adiabatic index $\gamma$, characterizing the equation of state of medium. This means that the form of the turbulence spectrum (6.17) does not depend on whether or not the pressure gradient term is taken into account in the Euler equation. As a result, the energy spectrum (6.17) describes also turbulence based the one-dimensional Hopf equation. Indeed, in (6.17), there is no dependence on the parameters $c_\infty$ and $\gamma$. This means the exact correspondence of the energy spectrum (6.17) to the solution of the Hopf equation, with which the solution (2.17) exactly coincides under the formal limit $c_\infty \to 0; \gamma \to 1$ [4]-[6]. Thus, for the solution of the Hopf equation, the energy spectrum (6.17) is obtained exactly in the limit $t \to t_0 (\gamma = 1)$, where $t_0(\gamma)$ is given in (4.2) [4]-[6].

Note that, taking into account the homogeneous dissipation, the representation for the energy spectrum under the condition $\mu < \mu_{th} = 1/t_0$ for the time tending to the minimum collapse time (5.3) $t \to \tilde{t}_0(\mu)$ has the form:

$$\tilde{E}(k) = C_E \left(1 - t_0 \mu\right)^2 k^{-8/3} \tag{6.18}$$

Note that despite the similarity of the spectra (6.17) and (6.18), there is a significant difference in the conditions of their implementation, when, in contrast to (6.17), the spectrum (6.18) is realized at the collapse time (5.3), which additionally depends on the level of dissipation in the system.

When taking into account the effective viscosity and using the exact solution (5.17) instead of the solution (2.17), the energy spectrum in the limit $t \to t_0$ has the form $E(k) = C_E k^{-8/3} \exp\left(-2t_0 k^2 \nu\right)$. Therefore, only in the inertial range of scales $L^{-1} << k << l_\nu^{-1} = \left(2t_0 \nu\right)^{-1/2}$ the spectrum of energy (6.17) is realized.

Thus, on the basis of exact solutions of one-dimensional Euler equations for compressible medium in the form of a simple wave, the universal spectra of energy, density and pressure with the index of -8/3 are obtained. The same spectrum is also relevant to the turbulence without pressure in the frame of the Hopf equation. In the spectrum (6.17), not only the exponent -8/3, but also the constant value $C_E$ for both the solution of the Hopf equation and the solution of the Euler equations exactly coincide with each other.

To obtain the spectrum (6.17), we used space averaging rather than statistical ensemble averaging. In this case, it turns out that for any initial energy spectrum, the limiting turbulence regime with the energy spectrum (6.17) is realized. For example, to the initial velocity field $V_0(x) = a \exp(-x^2 / 2x_0^2)$ the initial energy spectrum $E_0(k) = a^2 x_0 \exp(-x_0^2 k^2)$ is corresponded. Therefore, regardless of the type of the initial turbulence energy spectrum of simple waves, the universal turbulence energy spectrum (6.17) with the index -8/3 is realized during their nonlinear collapse.

## Spectrum of dissipation rate

Let us consider also the exact solution for the spectrum

$E_D(k;t) = \dfrac{1}{2\pi} \int\limits_{-\infty}^{\infty} dr R_D(r;t)\exp(-ikr); R_D = \dfrac{1}{L}\int\limits_{-\infty}^{\infty} dx \tilde{\varepsilon}(x+r;t)\tilde{\varepsilon}(x;t)$ of the energy dissipation rate fluctuations

$\tilde{\varepsilon}(x;t) = \dfrac{v}{2}\left(\dfrac{\partial\langle V(x;t)\rangle}{\partial x}\right)^2$, where we use exact regular at all times solution (5.17). The spectrum corresponding to

the representation (6.13) has the form similar to (6.15):

$$E_D(k,t) = \frac{1}{2\pi L}I_D(k)I_D(-k);$$

$$I_D(k) = \frac{v e^{-tk^2 v + iktc_\infty}}{2}\int\limits_{-\infty}^{\infty} dx \frac{dV_0}{dx}\exp\left(ikS(x;t)\right)I_S(x;t); \qquad (6.19)$$

$$S(x;t) = x + t\frac{(\gamma+1)}{2}V_0(x); I_S(x;t) = \frac{1}{2\sqrt{\pi t\, v}}\int\limits_{-\infty}^{\infty} dy \frac{dV_0}{dy}\exp\left(-\frac{(S(x;t)-S(y;t))^2}{4t\,v}\right)$$

In the limit $kx_M \gg 1$, we use the stationary phase method when evaluating the integral $I_D$ in the limit $t \to t_0$. Thus it is possible to obtain estimation $I_D \cong \dfrac{v}{2}V_0'(x_M)e^{-t_0 k^2 v + ik(c_\infty t_0 + S(x_M;t_0))}I_1(k)I_S(x_M;t_0)$, where $I_1(k) \propto O\left(k^{-1/3}\right)$ is represented in (6.16) when in the limit $v \to 0$, $t \to t_0$ here $I_S(x_M;t_0) \to V_0'(x_M)t_0/(t_0-t)$.

As a result, we obtain a limit representation for the spectrum of fluctuations in the rate of dissipation of turbulent

energy that is no longer dependent on the kinematic viscosity coefficient and has the form:

$$E_D(k) = C_D k^{-2/3}\exp\left(-t_0 k^2 v\right)$$

$$C_D = 2^{-1/3}L^{-1}v^2\left(\frac{dV_0}{dx}\right)_{x=x_M}^{8/3}\left(\frac{d^3 V_0}{dx^3}\right)_{x=x_M}^{-2/3}I_S^2(x_M;t_0)\Phi^2(0) \qquad (6.20)$$

The exponent of the spectrum (6.20) is consistent with the results of observations of small-scale intermittency of atmospheric turbulence [9] (see Figure 103 in [9] and discussion in the next section) [31] in the inertial range of scales $L^{-1} \ll k \ll l_v = (2t_0 v)^{-1/2}$. Indeed, spectrum $E_D \propto k^{-2/3}$ in (6.20) for this range of scales.

# 7. Discussion and comparison with observational data

We note some applications of the exact solutions (6.20) and (6.17) for the energy dissipation rate fluctuation spectrum and the turbulence energy spectrum, respectively.

The problem of strong interaction in developed turbulence, as in the physics of critical phenomena and elementary particles, is characterized primarily by the manifestation of the properties of small-scale fluctuations at all intermediate scales, up to macroscopic wavelengths [28].

In this regard, it is important to take into account fluctuations in the energy dissipation rate, which can serve as a measure of the observed small-scale intermittency of turbulence [9], [13]. For this purpose, the spectrum of fluctuations in the energy dissipation of atmospheric turbulence near the water surface (at a distance of 1 meter from the water surface) in the range of wave numbers corresponding to scales from 30 cm to 0.1 cm was obtained in experiment [29] (see also Figure 103 in [9]). The exponent of the spectrum was found to be equal to the value $0.62 \pm 0.05$ in [29]. Measurements involving a smaller-scale turbulence in the surface layer of the atmosphere (at an altitude of 13.5 meters from the earth's surface) than in [29] led to a refinement of the value of this spectrum indicator equal to $0.67 \pm 0.05$ [31] (see Figure 5 in [31]).

Thus, the exponent of the spectrum -2/3 of the solution (6.20) coincides with a good accuracy with the average value 0.67 of the exponent obtained for the spectrum of energy dissipation fluctuations in the experiments of Kholmyansky (1972) [31] (see Figure 5 in [31]).

Previously, a theoretical interpretation of similar experimental data [30] was given in [22].

Physically model in [22] means concentration of the dissipation in a small fraction of volume-asymptotically on a fractal set [17]. At the same time, in [22], for the correlation function and the spectrum of turbulence energy dissipation fluctuations, the representations $b_{EE} \propto r^{-\mu}$ and $E_{EE} \propto k^{-1+\mu}; 0 < \mu < 1$ (where $\mu = \dfrac{\ln \alpha}{\ln \beta}; \alpha = \lambda_1 / l_1 << 1; \beta = \lambda_2 / \lambda_1 << 1$), respectively are obtained. In this case, $\alpha$ - is the ratio of the size of the region in which the dissipation of turbulence energy is concentrated to the size of the region of the turbulent flow free from dissipation. The value $\beta$ characterizes the ratio of the size of the regions in which the energy dissipation is concentrated, when $\lambda_2$ it corresponds to the next stage of scale division after the division stage, when the size of the dissipation region was equal $\lambda_1$.

At the same time, in the model [22], the value $\mu$ is the only fitting parameter, which, however, cannot be determined within the framework of this model and must be determined from comparison with the data of experimental observations. From the comparison with the exact solution of the turbulence T-problem obtained in (6.20) and (6.13) for the correlation function of the turbulence energy dissipation rate, it follows that the value of this parameter should be equal to $\mu = 1/3$.

Let us consider the results obtained in (6.17) for the analysis of some known observational data and available estimates of the turbulence spectra, in which the effects of compressibility of the medium play an important role. In particular, the exact solution of the Euler equations and the problem of turbulence in a compressible medium, given above, allow us to refine the known results of the theoretical study of the turbulence spectra of nonlinear waves, the breaking of which leads to the formation of shock waves in strong acoustic turbulence [35]. Indeed, in [35], isothermal sound nonlinear waves are described by the potential (vortex-free) flow of a medium, provided that its thermal conductivity is very high and its viscosity can be neglected. The three-dimensional Euler equations with random force are used, and an additional assumption is made that the local isothermal speed of sound $c_T$ is a constant value.

The main result of [35] is that the solution of the 3-D Euler equations for a compressible medium is reduced to the solution of the 1-D Hopf equation (1.9) for the velocity amplitude $u = u_p$ of a wave packet propagating primarily in one selected direction (along the z-axis, for example) inside a narrow cone with an angle $\alpha_{cone} = k_\perp / k_z < \sqrt{u_p / c_T}$, when it is also possible to neglect the dispersion. On this basis, it is concluded in [35] that the breaking of simple waves $u_p$ and the formation of shock discontinuities must inevitably occur. Thus, it is shown in [35] that the mechanism of occurrence of strong acoustic turbulence is associated with the collapse of the solution of the 1-D Hopf equation.

Therefore, the turbulence energy spectrum $E(k) \propto k^{-8/3}$ in (6.17), corresponding precisely to the collapse of the solution of the Hopf equation, is an exact solution in the theory of strong acoustic turbulence within the framework of the formulation proposed by Kadomtsev and Petviashvili in [35]. However, in [35] it is proposed to use another well-known energy spectrum $E(k) \propto k^{-2}$ to describe acoustic turbulence which is heuristically obtained for a random set of shock waves (see also [36], [37]).

The spectrum of density or pressure pulsations in our theory will have the same power-law dependence on the wave number as in (6.17) or (6.18). In this regard, it is of interest to compare the turbulence spectrum (6.17) with the results of observations of turbulence spectra of electron density pulsations and the magnetic field pulsations in the magnetosheath and solar wind which are anti-correlated with each other [42]-[47]. Indeed, similar values of exponents close to -8/3 (see Fig. 1 and Fig.2 below, taken from [42] and [46] respectively) were obtained in the observation data of turbulence spectra for the solar wind and for the magnetosheath [42]-[47].

In [42] it is noticed that most of the energy of the turbulence structures cascades along the flow direction with a power law close to the observed 1-D spatial spectrum $E(k) \propto k^{-8/3}$ that is steeper than all those predicted by the existing magneto-hydrodynamic theories, which are based on oversimplifying assumptions such as isotropy and incompressibility which gives the spectrum $E(k) \propto k^{-7/3}$ [63] (see Fig.1). As also noticed in [42], spectrum $E(k) \propto k^{-8/3}$ does not break down below the ion scale $k_V \rho_i \geq 1$ ($\rho_i = 75 km$-the ion Larmor radius; $k_V$ - wave vector component along the flow direction, see Fig.1) as was usually thought concerning the physics of the mirror mode [48], [49]. Indeed, in the kinetic linear theory of the "mirror" instability in magnetosheath plasma, this instability is impossible below the ion scale and its absence needs a satisfactory nonlinear theory to explain the generation of the observed small scales [42]. Thus, from the observational data [42], [43] it is possible to conclude that the largest mirror structures (which have frequencies $f_L \approx 0.3 f_{cp} = (0.114 - 0.084) \sec^{-1}; T_L = f_L^{-1} = (8.8 - 11.9) \sec$; here $f_{cp} = (0.33 \pm 0.05) \sec^{-1}$ - proton cyclotron frequency) can be a pumping source of energy which cascades more preferentially along the flow direction up to $k_V \rho_i \approx 3.5$. In this cascade, the creation of energetic small scales is realized when the strong nonlinear effects must overcome the linear damping [42].

While comparing the energy spectrum $E(k) \propto k^{-8/3}$ from (6.17) with the data of [42], it turns out that as a nonlinear mechanism shaping the observed energy cascade, this process can be a non-linear collapse of the mirror modes modeled using the simple slow wave corresponding to the minus sign in the Hopf equation (1.9).

Indeed, the mirror mode is propagating very slowly in plasma [42], [48], [49]. Let us use in (4.4) the characteristic values for the velocity and the scale of the flow length $x_0 \approx 2000$ km and $a = 248 \pm 25$ km/sec [42], [43]. At the same time, for the case of plasma the adiabatic index $\gamma = 5/3$ and the estimate of the time of the collapse of a nonlinear wave based on (4.4) give a value $t_0 = (9.1 - 11.1) \sec$ that is close to the above value $T_L = (8.8 - 11.9) \sec$ [42], [43].

The energy spectrum (6.17) corresponds also to the Hopf equation describing the motion of particles by inertia. In this connection, let us consider the question of under what conditions the cosmic plasma particles can actually move by inertia with zero total acceleration.

For the plasma particles moving perpendicular to the direction of the background magnetic field $B_{0z} = const$ this condition has the form (if in equation (2) in [64] left part (2) equal zero) $\dfrac{l_{B'}}{l_{p'}} = \dfrac{2C}{\beta_\perp}$.

Where $C = B'_z / B_{0z}$ is the magnetic field compression, $B'_z$ is the longitudinal magnetic field perturbations, the parameter $\beta_\perp$ is the ratio of the pressure perturbations $p'_\perp$ to the magnetic energy density associated with $B_{0z}$, $l_{B'}$ and $l_{p'}$ are the characteristic scales of $B'_z$ and $p'_\perp$ respectively. It is assumed that the transverse perturbations of the magnetic field can be neglected, which is typical for mirror modes [42], [48], [49] and that the characteristic time of the parameters $p'_\perp$ $B'_z$ are greater (see [47]) than gyro-period of the ions $f_{cp}^{-1} \approx 3 \sec$ [42]. We take into account that for a simple wave $l_{\rho'} = l_{\rho'} / \gamma$, where $l_{\rho'} = \left( \dfrac{1}{\rho} \left| \dfrac{\partial \tilde{\rho}}{\partial x} \right| \right)^{-1}$ - is the characteristic scale of the change in the

perturbation density, for which we obtain an estimate from (5.15) (assuming $t_{00}\mu \cong 1; t \cong t_{00}$)

$$l_{\rho'} = \frac{x_0}{M\sqrt{e}}\left(1 - \frac{(\gamma-1)M}{2\sqrt{e}}\right)^{\frac{3-\gamma}{\gamma-1}}; M = \frac{a}{c_\infty}.$$

Let the scale of longitudinal magnetic field fluctuations $l_{B'} \approx \lambda_B$ be equal to the magnitude of the screening radius

$\lambda_B$ of the ions [49], which describes the characteristics of magnetic field distribution in mirror modes. At the same

time, according to the observations [49], we can use the estimate $\lambda_B k_\perp = 1/\sqrt{\beta_\perp}$ where $k_\perp \approx 1/x_0$. As a result, the condition of the motion by inertia takes the form $\dfrac{M\gamma\sqrt{e}}{\left(1 - \dfrac{M(\gamma-1)}{2\sqrt{e}}\right)^{\frac{3-\gamma}{\gamma-1}}} = \dfrac{2C}{\sqrt{\beta_\perp}}$.

For example, (see Table 1 in [46]) in the case of the solar wind $C = 0.3; \beta_\perp = 0.4; M = 0.3$ and $\gamma = 5/3$, when the left side of that condition is equal to 0.934, and the right side is equal to 0.949, which indicates an accuracy of about 1.5% for the condition whereby the inertial motion of the plasma particles is indeed realized.

Therefore, based on the obtained estimates, it can be concluded that to describe the motion of particles in the solar wind, it is permissible to use the Hopf equation, but for the magnetosheath it is need to use the Euler equations and its solution in the form of the simple wave.

Indeed, the universal spectrum (6.17) and (6.18) with an index of -8/3 is in agreement with the observational data for both the solar wind and the magnetosheath. Moreover, as can be seen from Fig. 2, for the magnetosphere, the region of realization of the spectral law -8/3 is noticeably larger than the similar region observed for the solar wind. From the observations in the Saturn magnetosphere [45], it follows that there is only a spectrum close to the -

8/3 law is observed in the entire inertial range of scales immediately following the energy pumping region. To explain this fact, it was suggested in [45] that the cascade mode corresponding to the -5/3 law does not have time to be established during the passage of solar wind particles through the magnetosphere from the bow shock wave to the magnetopause.

In this regard, the question of the conditions under which the -8/3 law can be close to the real energy spectrum in the inertial range of scales for a turbulent regime characterized by a finite integral energy is relevant. And vice versa, when should the regime corresponding to the -5/3 law prevail? In [33], as already noted in the Introduction, the law-8/3 was probably introduced for the first time, not only as a limiting law allowing energy dissipation in the absence of viscosity, but also its representation was obtained on the basis of dimensional considerations when using integration over the entire space instead of statistical averaging.

To answer these questions, we obtain estimates of the integral time of the entire cascade process of energy transfer, from the external scale $L$ of turbulence to the extremely small scale for each of the turbulent modes corresponding to the laws of $-5/3$ and $-8/3$. In fact, this allows us to estimate the relative durations of each of these modes. The longer this relative duration is for a given mode, the more significant the contribution of this mode to the resulting description of the turbulent flow can be.

We use a modification of the formula proposed by Onzager to estimate the implementation time of the cascade process in the -5/3 law [17] (see formula (36) in [17]). It has the form $T_h = \int\limits_{1/L}^{\infty} \frac{dk}{k} \tau_h(k)$, where $\tau_h(k) \cong \frac{1}{C_h k^{-(1-h)}}$ is the energy transfer time for a given value of the wave number, and the parameter $h$ (the Holder parameter) characterizes the smoothness of the turbulent velocity field $\Delta V = |V(x+r) - V(x)| \cong C_h r^h$. Indeed, the characteristic time for motion with a wavelength $r \propto 2\pi/k$ has the form $\tau_h \propto \frac{r}{\Delta V} = \frac{r^{1-h}}{C_h}$ (see also a similar estimate in [57]). In this case, the implementation time of the turbulent regime corresponding to the -5/3 law $h = 1/3$, for which, has the form $T_{1/3} \propto L^{2/3}/C_{1/3}$ where $C_{1/3} = \frac{2}{3} C_K^{1/2} \langle \varepsilon \rangle^{1/3}$ and the Kolmogorov constant $C_K \approx 0.5$ [9]. The spectrum (6.17) corresponds to a structural function $S_2 \propto r^{5/3}$, which corresponds to the value of the parameter $h = 5/6$. At the same time, the time of implementation of the turbulent regime with the law -8/3 has the form $T_{5/6} \propto L^{1/6}/C_{5/6}$, where $C_{5/6} = \frac{1}{6}C_E^{1/2}$ and $C_E$ is defined in (6.17). Note that the implementation time of this mode is much longer than the singularity time of the solution defined in (4.2) $T_{5/6} >> t_0$. In particular, for the example of the initial velocity field discussed in (4.4) and further on we have relation $T_{5/6} \approx 20 t_{00}$.

As a result, the ratio of these times is equal to $T_{1/3}/T_{5/6} \propto L^{1/2}C_E^{1/2}/4C_K^{1/2}\langle\varepsilon\rangle^{1/3} \equiv \chi$. For example, for the initial velocity field $V_0(x) = \frac{\overline{V}_0}{\sqrt{2\pi}}\exp(-2\pi x^2/L^2); \overline{V}_0 = const$ corresponding to the example in (4.4), we obtain

$\chi \approx 0.116 \frac{\overline{V}_0}{|\Delta V_L|}; |\Delta V_L| = (\langle\varepsilon\rangle L)^{1/3}$. Where $\overline{V}_0$ - is the average velocity corresponding to the given initial distribution of the velocity field and $|\Delta V_L|$ -is the change in the velocity of turbulent motion over a distance $L$ (see also (33.6) in [13]). Here $L$ corresponds to the external integral scale of turbulence defined in (6.2) and characterizes the main energy mode. For example, according to [46] (see Table 1 and Table 2) for the solar wind, we obtain an estimate $\chi_{SW} \approx 3.24$, that corresponds to the values $L \approx 2.356 \times 10^9$ m; $\overline{V}_0 \approx 330 \times 10^3$ m/sec; $\langle\varepsilon\rangle \approx 700$ m$^2$/sec$^3$. Similarly, for the magnetosheath, we obtain $\chi_{MSH} \approx 0.414$, where the values $L \approx 1.76 \times 10^8$ m, $\overline{V}_0 \approx 275 \times 10^3$ m and $\langle\varepsilon\rangle \approx 2.6 \times 10^6$ m$^2$/sec$^3$ are used.

Thus, these quantitative estimates are consistent with the qualitative conclusions made in [45] and indicate that not only in the Saturn magnetosphere, but also in the Earth's magnetosphere, according to the data used above from [46], the implementation of the spectral law -8/3 of Sulem-Frisch [32] dominates in comparison with the manifestation of the law -5/3 of Kolmogorov-Obukhov.

For comparison, we also give an estimate of the parameter $\chi_{SL} \approx 0.97$, which follows from observations of small-scale intermittency in the turbulence of the surface layer of the atmosphere of Kholmyansky (1972) [31]. This estimate corresponds to the value of the average velocity $\overline{V}_0 \approx 5.7$ m/sec and $L \approx 31.25$ m is the Monin-Obukhov scale is chosen as the length scale for the average dissipation rate $\langle \varepsilon \rangle \approx 0.01$ m$^2$/sec$^3$ [31]. At the same time, the spectrum of turbulence energy, according to observations [31] (see Fig.4 in [31]), corresponds only to the Kolmogorov-Obukhov law -5/3 in the scale range available for measurement. Apparently, the lack of the possibility of reliable measurements in a wide range of rather small scales, where, however, viscous dissipation is not yet significant, did not allow in [31] to reveal the presence of the law-8/3, as is the case in the above-considered observational data in the cosmic plasma.

Indeed, this is also indicated by the noted good correspondence between the spectrum (6.20) and the spectrum of energy dissipation rate fluctuations obtained in [31], which refers to a much wider range of small scales of the inertial interval (see Figure 5 in [31]) than for the energy spectrum (see Figure 4 in [31])

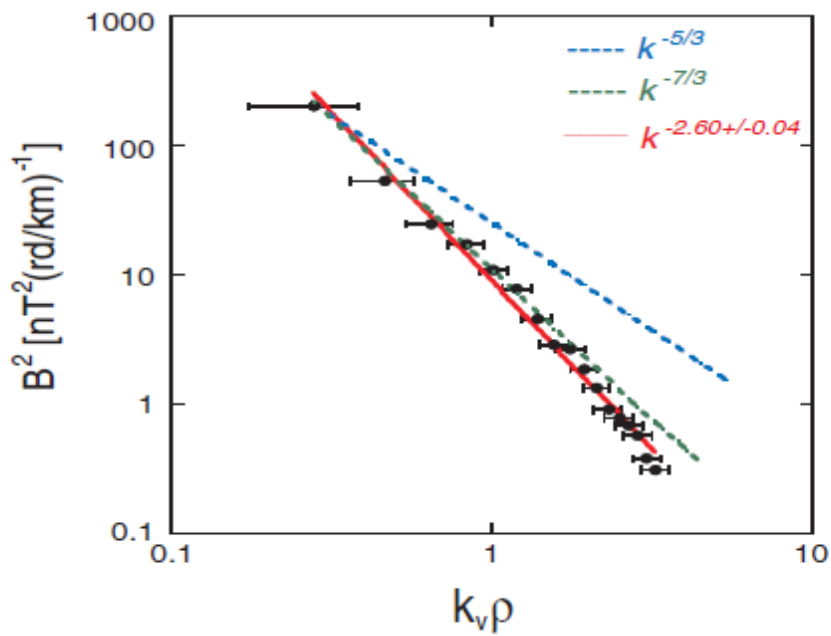

Fig. 1

Magnetic field turbulence spectrum for the magnetosheath (when proton Larmor radius $\rho = 75$ km). The red line is a direct fit revealing a power law $k^{-2.6}$. Two other power laws are plotted for comparison: $k^{-7/3}$ (green) and $k^{-5/3}$ (blue) [42] (see Fig.6 in [42]).

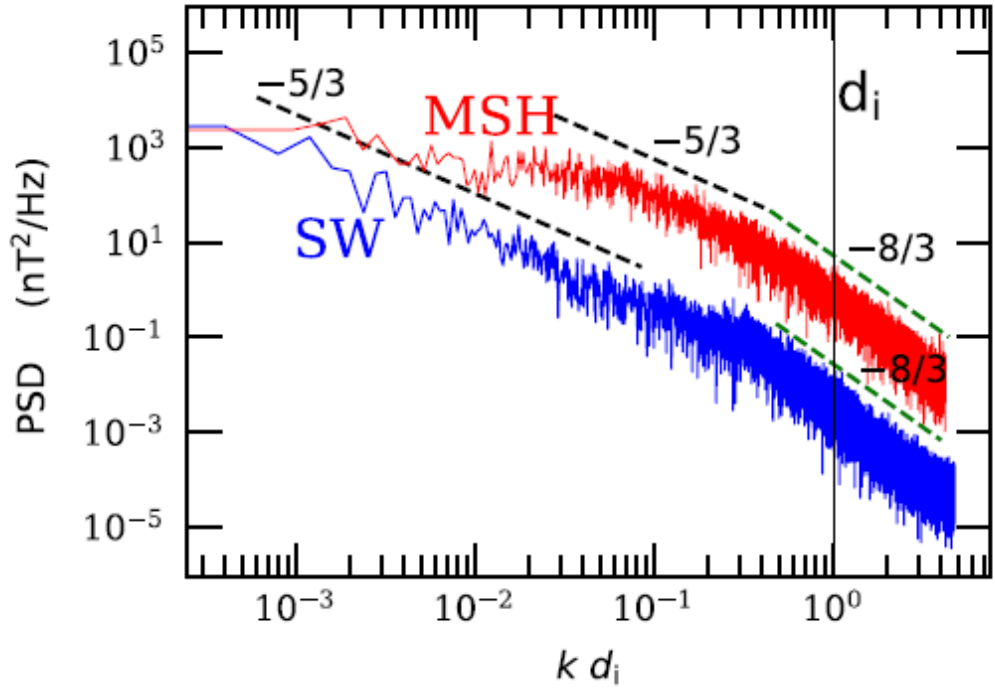

Fig.2

Magnetic field turbulence spectra for the solar-wind (SW) in blue (when ion-inertial length $d_i = 75$ km; $\left\|\langle \vec{V} \rangle\right\| = 330$ km s$^{-1}$ the speed of SW) and magnetosheath (MSH) in red (when $d_i = 56$ km; $\left\|\langle \vec{V} \rangle\right\| = 278$ km s$^{-1}$ ) interval. The solid vertical line represents $kd_i = 1$ with the wave vector $k = 2\pi f \, / \left\|\langle \vec{V} \rangle\right\|$, where $f$ is the frequency [46] (see Fig.1 in [46]).

## Conclusions

Thus, the exact closed-form explicit analytical solution to the Riemann problem for the Euler one-dimensional hydrodynamics equations is obtained in two different forms. The first of these forms corresponds to a simple Riemann wave. The second more general form of the exact solution describes the isentropic flow, for which analytical representations are obtained explicitly for the Riemann invariants at a certain value of the adiabatic exponent $\gamma = 3$. It is possible to use both forms of exact analytical solution for validation of well-known numerical

Godunov-type methods [16] which are used in hydrodynamics and gas dynamics research on the base of the Riemann solver.

The regularization for that solution on unlimited time is obtained when dissipation is taken into account. That gives an unexpected positive resolution for the generalization of the Clay problem [11] in the compressible case.

The explicit representation for the shock wave arising time is obtained for arbitrary initial conditions of a simple wave.

Closed explicit analytical representations are obtained for one-point and two-point moments of hydrodynamic fields and for the energy spectrum, which gives an example of solving the main turbulence problem based on the exact solution of one-dimensional Euler equations for a compressible medium. In particular, exact solutions for the structural functions of different orders and for the finite value of the average rate of dissipation of the turbulence energy in the zero-viscosity limit in the connection with Onsager's dissipative anomaly are obtained. It is shown that the scale invariance of the turbulent regime inherent in the original Euler equations is preserved in the entire region of regularity of the solution and is violated only in the region of its collapse corresponding to the moment of collapse of a nonlinear simple wave.

An exact solution for the spectrum $E_D \propto k^{-2/3}$ of fluctuations in the turbulence energy dissipation rate is obtained in a good agreement with the observations [31] of the effects of intermittent turbulence in the surface layer of the atmosphere.

The universal turbulence spectrum power-law $E(k) \propto k^{-8/3}$, obtained from the exact solutions of the Euler and Hopf equations, corresponds to the known Kadomtsev-Petviashvili acoustic turbulence theory [35]. Also the turbulent spectra of the same form $k^{-8/3}$ are observed in the Earth's and Saturn's magnetospheres, as well as for the solar wind [42]-[47]. We propose a mechanism for the sink of the turbulent energy which has not been considered earlier and which is caused by the collapse of nonlinear simple waves in the form of the mirror modes. This gives a new base for understanding the known problem as to how collision-less plasmas energy dissipate [46].

The scaling law presented here could be used as a new input for the turbulent reconnection models, which aim to explain fast reconnection in astrophysical plasmas by considering large- to small-scale energy transfers. It can also be considered as an external constraint in numerical models devoted to study the reconnection driving problem [65].

The well-known property of inertia for all material objects is the most obvious and inexhaustible mystery in nature. This was clearly understood by L. Euler, E. Mach, A. Einstein and others (see [66]). In our paper, a new universal aspect of inertia motion is identified, which is related to the dynamics of a continuous medium in the theory of hydrodynamic turbulence. Indeed, until now it is generally accepted that without taking into account the pressure in the equations of hydrodynamics, it is impossible to construct an adequate reality of the theory of turbulence. There is even a humorous statement about this by a well-known theoretical physicist who compared the theory of turbulence without pressure with a someone who has lost his manhood. The result obtained in this paper, however, is a counterexample to this "theorem", at least for the case of turbulence in a compressible medium.

Indeed, in our paper it is established that there is an exact coincidence of the universal spectrum of turbulence energy obtained on the basis of the exact solution of the 1-D Hopf equation describing the motion of fluid particles by inertia, and the energy spectrum obtained on the basis of the exact solution of the 1-D Euler equations in the form of a simple nonlinear wave, when the pressure is taken into account. This coincidence is due to the consideration of the both solutions in the area of the collapse of a nonlinear wave and the occurrence of a corresponding singularity, which leads to a universal form of the energy spectrum when a shock wave occurs. This conclusion is important for understanding the conditions for application obtained in [4]-[6] for the exact solution of the problem of potential (acoustic) and vortex turbulence in a compressible medium which corresponds to the exact solution in Euler variables for the multi-dimensional Hopf equation.

For example, such an equation reduces the description of clustering processes in the physics of granular matter [67] (see Equation (9) in [67]) and the corresponding processes in astrophysics in solving the problem of the formation of the large-scale structure of the universe [68] (see Equation (7.6) in [68]).

Earlier, attention was repeatedly drawn to the analogy between the manifestations of scale invariance and universality of the energy spectrum indicators in developed turbulence and in the description of critical phenomena [28]. The conclusion obtained in our work clarifies the scope of this analogy. Indeed, the strong interaction between movements of different scales in the developed turbulence corresponds to the appearance of long-range correlations between all modes at the critical point of the phase transition, for which scale invariance is realized. However, the scale invariance is known to be violated everywhere outside the critical point [28]. On the contrary, for strong turbulence, the scale invariance characteristic of the Euler equations takes place during the entire time of the evolution of the solution of these equations and is violate only at the moment of the collapse of the solution, as shown by the example of the scale invariance of the kurtosis.

On the other hand, in the region of the singularity of the solution, when ordered coherent structures are formed in the form of shock waves, the symmetry of the system decreases. In this case, there is a similarity, in particular, with phase transitions of the second kind, which are characterized by a decrease in the symmetry of the system at the same energy state [52]. There is also an analogy with the change of continuous symmetry of the gauge type, characteristic of the adiabatic invariant of a linear oscillator with a time-variable frequency, to discrete symmetry (such as mirror reflection or rotation by an angle of 180 degrees) in the region of parametric frequency resonance, where the preservation of the adiabatic invariant is violated [69]-[71].

As is known from the physics of shock waves [13], [14], a positive jump in entropy is always formed behind the shock wave front, which actually compensates for the decrease in entropy that is associated with the formation of an ordered coherent structure of the shock wave front itself and the above-mentioned decrease in symmetry. Therefore, a shock wave propagating at a speed exceeding the maximum velocity of perturbation propagation in the medium can be considered as an object for which the factors of symmetry and information are dominant.

Thus, the conservation laws associated with scale-invariant symmetry can play an important role for a wide class of nonlinear systems and phenomena, for which factors of a symmetry and information may be dominant in comparison with space-time and energy characteristics.[28], [66], [72]-[75].

We thank Ya. E. Krasik for the permanent support, E. A. Novikov for his positive estimation of our work and also Ya. G. Sinai and V. E. Fortov for attention and stimulating discussions on their seminars in July and October 2019 in Moscow. We are grateful to E. Bormashenko, G. Fal'kovich and E. A. Kuznetsov for their useful comments and to M. Z. Kholmyansky for discussion on his experimental data, obtained under the supervision of A. S. Gurvich and published in [31].

The study is supported by the Russian Science Foundation, Grant number: 14-17-00806P and by the Israel Science Foundation, Grant number: 492/18

Data Availability Statements: "The data that support the findings of this study are available from the corresponding author upon reasonable request.